\documentclass[aps,prl,nofootinbib,superscriptaddress,letterpaper,amsmath,amssymb]{revtex4}
\pdfoutput=1

\usepackage{graphicx}  
\usepackage{dcolumn}   
\usepackage{bbm}        
\usepackage{amsmath}
\usepackage{amsfonts}
\usepackage{amssymb}   
\usepackage{epstopdf}
\usepackage{slashed}
\usepackage{multirow}
\usepackage{color}
\usepackage{float}

\usepackage{verbatim}

\def\gsim{\lower0.5ex\hbox{$\:\buildrel >\over\sim\:$}}
\def\lsim{\lower0.5ex\hbox{$\:\buildrel <\over\sim\:$}}

\newcommand{\be}{\begin{equation}}
\newcommand{\ee}{\end{equation}}
\newcommand{\bea}{\begin{eqnarray}}
\newcommand{\eea}{\end{eqnarray}}

\newcommand{\nbox}{{\,\lower0.9pt\vbox{\hrule \hbox{\vrule height 0.2 cm
\hskip 0.2 cm \vrule height 0.2 cm}\hrule}\,}}
 \newcommand{\vev}[1]{\langle {#1} \rangle}
 \newcommand{\lagr}{\mathcal{L}}

\def\sub#1{_{\lower.25ex\hbox{$\scriptstyle#1$}}}

\newskip\zatskip \zatskip=0pt plus0pt minus0pt
\def\matth{\mathsurround=0pt}
\def\lsim{\mathrel{\mathpalette\atversim<}}
\def\gsim{\mathrel{\mathpalette\atversim>}}
\def\sigv{\ifmmode \langle\sigma v\rangle\else $\langle\sigma v\rangle$\fi}
\newskip\zatskip \zatskip=0pt plus0pt minus0pt
\def\matth{\mathsurround=0pt}
\def\lsim{\mathrel{\mathpalette\atversim<}}
\def\gsim{\mathrel{\mathpalette\atversim>}}
\def\atversim#1#2{\lower0.7ex\vbox{\baselineskip\zatskip\lineskip\zatskip
  \lineskiplimit
  0pt\ialign{$\matth#1\hfil##\hfil$\crcr#2\crcr\sim\crcr}}}

\begin{document}

\thispagestyle{empty}
\vspace*{-3.5cm}

\vspace{0.5in}

\title{Indirect Detection Constraints on s and t Channel Simplified Models of Dark Matter }

\begin{center}
\begin{abstract}
Recent Fermi-LAT observations of dwarf spheroidal galaxies in the Milky Way have placed strong limits on the gamma-ray flux from dark matter annihilation. In order to produce the strongest limit on the dark matter annihilation cross-section, the observations of each dwarf galaxy have typically been ``stacked" in a joint-likelihood analysis, utilizing optical observations to constrain the dark matter density profile in each dwarf. These limits have typically been computed only for singular annihilation final states, such as $b\bar{b}$ or $\tau^+\tau^-$. In this paper, we generalize this approach by producing an independent joint-likelihood analysis to set constraints on models where the dark matter particle annihilates to multiple final state fermions. We interpret these results in the context of the most popular simplified models, including those with s- and t-channel dark matter annihilation through scalar and vector mediators. We present our results as constraints on the minimum dark matter mass and the mediator sector parameters. Additionally, we compare our simplified model results to those of Effective Field Theory contact interactions in the high-mass limit.
\end{abstract}
\end{center}

\author{Linda M. Carpenter}
\affiliation{The Ohio State University, Columbus, OH}
\author{Russell Colburn}
\affiliation{The Ohio State University, Columbus, OH}
\author{Jessica Goodman}
\affiliation{The Ohio State University, Columbus, OH}
\author{Tim Linden}
\affiliation{The Ohio State University, Center for Cosmology and AstroParticle Physcis (CCAPP), Columbus,
OH 43210}

\maketitle


\section{Introduction}

Multiple astrophysical observations provide extremely strong evidence indicating the existence of a dark matter (DM) particle. However, these observations give little indication of how (or whether) DM may interact non-gravitationally with the Standard Model (SM), and vast regions of parameter space exist that may successfully explain interactions between DM and the SM. These interactions are currently probed by three classes of experiments: those that directly probe the scattering of dark matter with SM nuclei, those studying the production of DM in colliders, and those indirectly studying the annihilation of DM into stable SM final states. Any given model of DM may produce signals in more than one class of experiment, and observations and correlations between these experiments will reveal significant insight into the parameter space of viable dark matter models.


A significant fraction of UV complete DM models demand that DM annihilate into multiple sets of SM particles. Fixed relations between various coupling constants in the theory may follow from symmetry considerations, such as gauge invariance or supersymmetric relations. In the context of indirect detection signatures for DM, we may expect to see a $\gamma$-ray spectrum formed from the composition of several annihilation final states --- a scenario which typically results in a relatively smooth $\gamma$-ray spectrum. In this work, we will focus our analysis on a set of models where DM annihilates to SM fermions in the final state.  These models have become popular to explain spectral features in the Galactic Center~\cite{Daylan:2014rsa} and are popular portals in collider searches for dark matter~\cite{Abercrombie:2015wmb}. To make our constraints generic, we do not present exclusions assuming a 100\% annihilation rate into a single SM channel, but consider composite DM annihilation spectra where the total annihilation rate among all fermionic channels is held constant. We compare the resulting dark matter annihilation spectrum and intensity against the stacked population of dwarf spheroidal galaxies \cite{Ackermann:2013yva}, utilizing a joint-likelihood analysis to determine the minimum allowable dark matter mass that can annihilate at the thermal annihilation cross-section while remaining consistent with the gamma-ray data.



One well-studied method of comparing or translating bounds from different search strategies is through the use of effective field theories (EFTs) \cite{Goodman:2010yf, Goodman:2010ku, Goodman:2010qn}. In EFT models, one assumes that the only accessible new degree of freedom is the DM itself and that any particles mediating the DM-SM interaction can be consistently integrated out, leaving behind a set of operators representing effective interactions. This offers a model-independent method of analyzing and comparing results from various DM searches while still capturing most of the kinematic features of the process being explored.  Various EFT scenarios have been used to probe collider scenarios with DM couplings to light quarks and gluons \cite{Aad:2015zva}, Higgs bosons \cite{Chen:2013gya, Carpenter:2013xra}, and electroweak gauge bosons \cite{Crivellin:2015wva, Chen:2013gya, Rajaraman:2012fu, Nelson:2013pqa, Lopez:2014qja}.  However, the limitations of EFTs are becoming apparent. The predictive power of EFTs becomes questionable once the effective cutoff and the mass of the mediating particles becomes smaller than the center of mass energy of the process~\cite{Papucci:2014iwa, Busoni:2013lha, Busoni:2014sya, Abdallah:2015ter}. Fortunately, in the case of indirect detection, the center of mass energy is generally much lower than in the case of collider-production of dark matter. We therefore expect the EFT analysis to be valid so long as the effective cutoff scale exceeds twice the dark matter mass. 

In cases where the EFT model fails, one can turn to simplified models where a new sector is added to mediate interactions between the SM and DM. The particle content of the least-complex simplified models consists of adding in a single mediating particle that communicates between DM and the SM at tree level.  We evaluate the constraints provided by the Fermi-LAT dwarf analysis on a group of simplified models including s-channel annihilation through a heavy vector boson as well as t-channel annihilation through a scaler mediator. In each case, we show that the simplified models match the EFT predictions for DM annihilations in the heavy mediator limit. This provides a basic test indicating where the EFT breaks down in DM annihilation processes.


The format of this paper is as follows. In Section~II we will employ current observations of the $\gamma$-ray flux from the population of dwarf spheroidal galaxies and describe the joint-likelihood algorithm used to constrain the dark matter annihilation parameter space. In Section~III, we produce generic lower-limits on the dark matter mass in models where the DM is allowed to annihilate to multiple SM fermion final states.  These mass bounds are then interpreted as effective coupling bounds in Section~IV, assuming an effective model.  In Section~V we will present parameter space exclusions on models mediated by both vector and scalar particles, and compare these limits against EFT models with contact interactions. In Section VI we will discuss the implications of our analysis and conclude.


\section{Indirect Detection from Dwarf Spheroidal Galaxies}
\label{sec:IDdwarfs}
The dwarf spheroidal galaxies (dSphs) observed in the Milky Way provide some of the tightest constraints on the $\gamma$-ray flux from DM annihilation due to a combination of their: (i) high DM densities, (ii) relative proximity, and (iii) insignificant astrophysical $\gamma$-ray backgrounds~\cite{Mateo:1998wg,McConnachie:2012vd}. Observations by both the Fermi-LAT collaboration and several sets of external authors~\cite{Abdo:2010ex, GeringerSameth:2011iw, Ackermann:2013yva, Geringer-Sameth:2014qqa, Ackermann:2015zua, Drlica-Wagner:2015xua, Geringer-Sameth:2015lua, Hooper:2015ula} have examined the population of dSphs and have found no convincing (5$\sigma$) evidence for a $\gamma$-ray excess coincident with the population of these sources. Using this information, these groups have set strong limits on the annihilation rate of DM as a function of its mass and final interaction states.

The photon flux $(\text{photons cm}^{-2} \text{ s}^{-1})$ at Earth expected from DM annihilations in the region of interest is given by:
\begin{equation}
\Phi_{\gamma}=\frac{1}{4\pi}
\sum_{\substack{f}}
\frac{\vev{\sigma v}_{f}}{2m_{\chi}^{2}}\int_{E_{\text{min}}}^{E_{\text{max}}}\left(\frac{dN_{\gamma}}{dE_{\gamma}}\right)_{f}dE_{\gamma} J,
\label{eq:gflux}
\end{equation}
where the J-factor $(\text{GeV}^2\text{cm}^{-5})$ is the line of sight integral of the DM density $\rho$, integrated over a solid angle: $\Delta\Omega$,
\begin{equation}
J=\int_{\Delta \Omega}\int_{l.o.s}\rho^{2}(\mathbf{r})dl d\Omega^{\prime}.
\end{equation}

\begin{figure}[tbh-]
\centering
\includegraphics[scale=1.0]{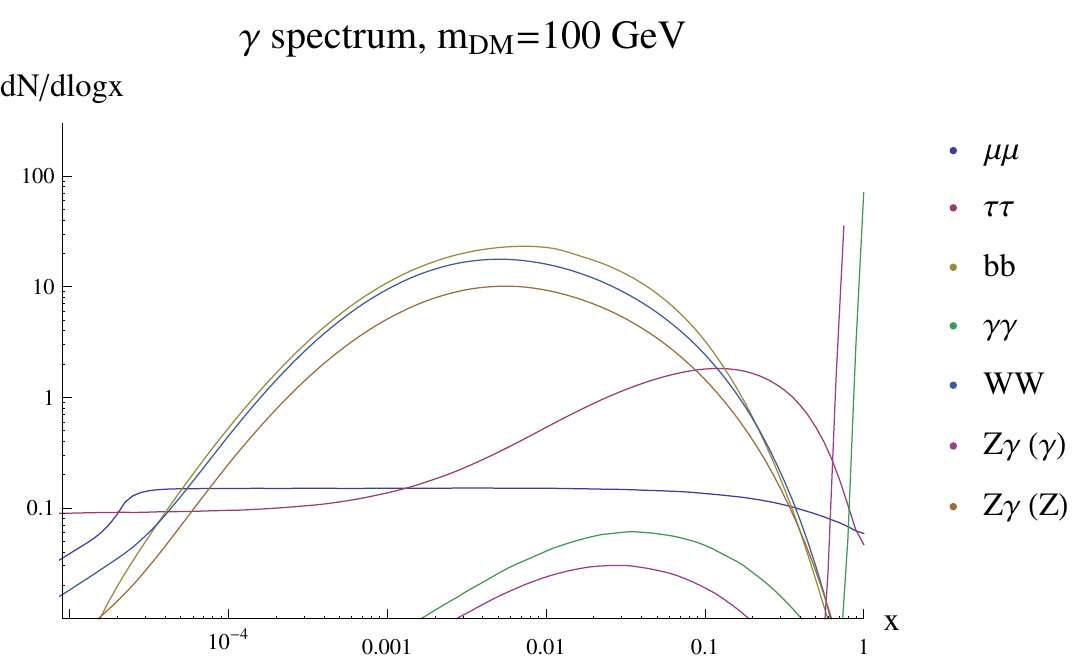}
\caption{Gamma ray spectrum from DM annihilation into various channels for a $100$ GeV DM particle.}
\label{fig:DMSpectrum100}
\end{figure}

\noindent The additional terms in Eq. \ref{eq:gflux} are dependent on the particle properties of the DM. Here, $dN_f/dE$ is the differential photon energy spectrum per annihilation to final state f, $\vev{\sigma v}_f$ is the thermally averaged DM annihilation cross section and $m_\chi$ is the DM mass. Throughout this work, we calculate the expected $\vev{\sigma v}_{f}$ for each kinematically accessible annihilation channel, and obtain the differential gamma ray spectra for annihilation channels, both from the Mathematica code PPPC 4 DM ID \cite{Cirelli:2010xx,Ciafaloni:2010ti} and from DarkSUSY \cite{Gondolo:2004sc}. In Figure~\ref{fig:DMSpectrum100} we show the resulting $\gamma$-ray spectrum for several choices of relevant SM pair annihilation channels assuming a 100~GeV DM particle. Here, the dimensionless parameter $x=E_{\gamma}/m_{\chi}$ is the gamma ray energy scaled by the DM mass.

\subsection{$\gamma$-ray Analyses of the Population of Dwarf Spheroidal Galaxies}

Most notably for this analysis, the Fermi-LAT collaboration has recently produced limits utilizing 6-years of Fermi-LAT data processed through the updated Pass 8 event reconstruction algorithm~\citep{Ackermann:2015zua}. In addition to performing an analysis of 25 individual dSphs, this analysis (like several before it) produces a joint-likelihood (or ``stacked") analysis of 15 dSphs with high confidence J-factors and background models. The analysis proceeds in the following way. The Fermi-LAT data is divided into 24 energy bins. For each energy bin and at the location of each dSph, a source with an extended $\gamma$-ray emission profile fit to observations is placed and the log-likelihood fit, LG($\mathcal{L}$), of this source is computed as a function of the source normalization in the energy bin. The $\Delta$LG($\mathcal{L}$) of this fit is computed by comparing this model against a background model (with no dSph) to the $\gamma$-ray data.
The Fermi-LAT team has publicly released these likelihood profiles for each dwarf \footnote{https://www-glast.stanford.edu/pub\_data/1048/}.

Noting that the WIMP paradigm mandates that the DM cross-section and spectrum be equivalent in all dSphs, the joint-likelihood analysis for a given DM model is computed as follows. For each dwarf the flux in each energy bin is computed by multiplying the expected DM flux by the astrophysical J-factor in that dwarf galaxy, and then the corresponding $\Delta$LG($\mathcal{L})$ is calculated, compared to the null hypothesis that dark matter does not annihilate. Since the J-factors of individual dwarf galaxies are highly uncertain, the astrophysical J-factor is allowed to shift from its measured value (J$_{meas}$) to its best fit value (J$_{bf}$), incurring a log-likelihood cost of $\Delta$LG($\mathcal{L})$~=~(J$_{bf}$ - J$_{meas}$)$^2$ / (2$\sigma_J^2$). The 95\% upper-limits computed by the Fermi-LAT collaboration are set when the total $\Delta$LG($\mathcal{L})$ for all dwarf-galaxies exceeds the model with no DM annihilation by 2.71/2.

\begin{figure}[H]
\centering
\includegraphics[scale=.9]{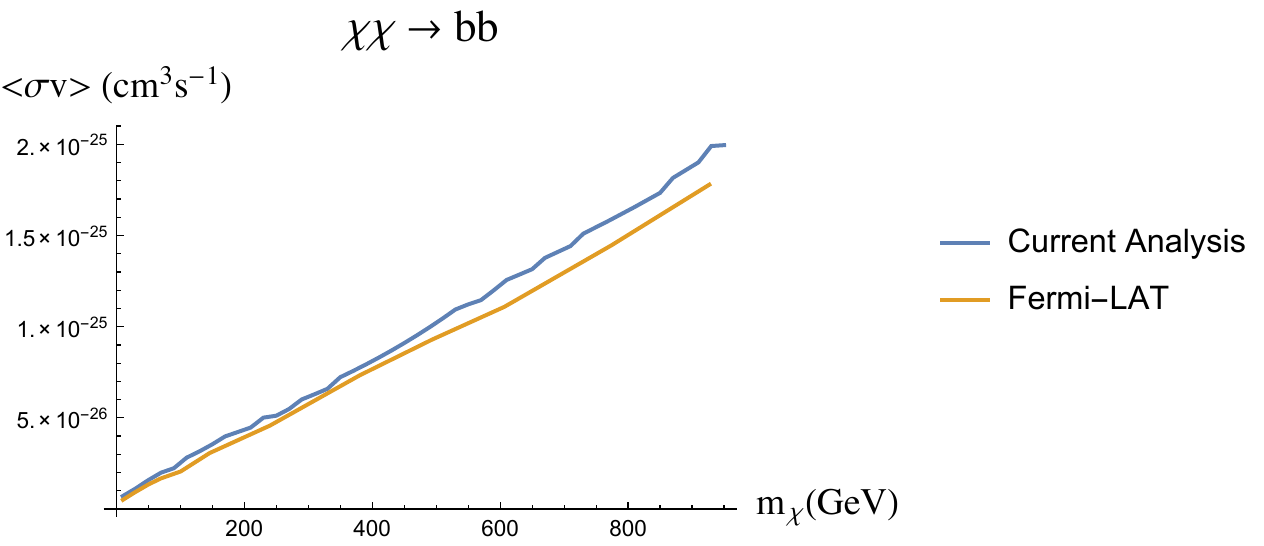}
\includegraphics[scale=.9]{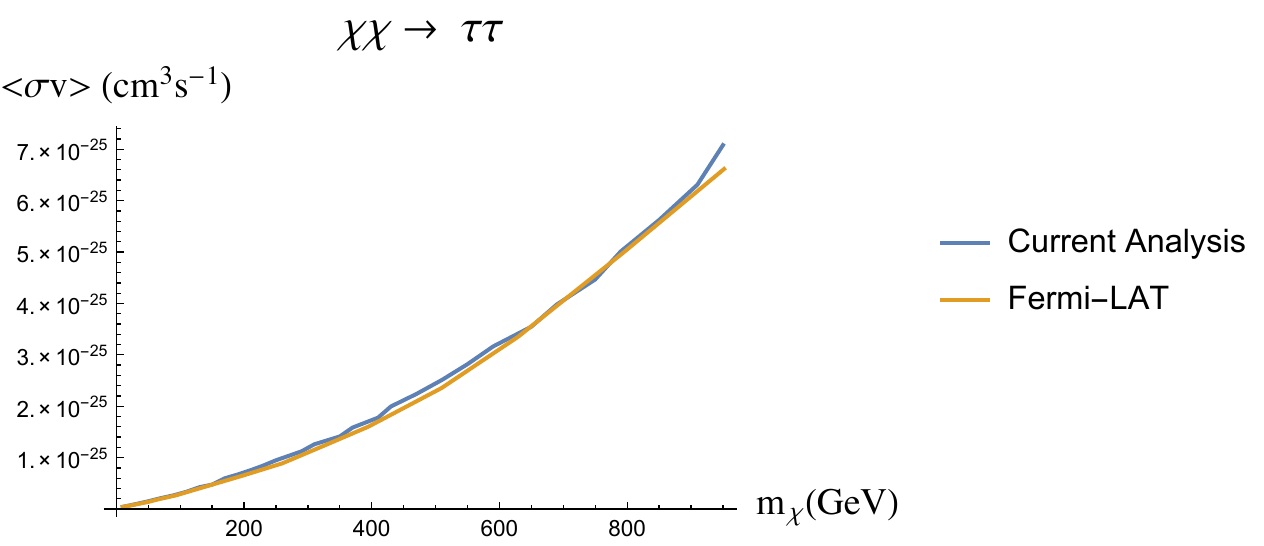}
\caption{A comparison of the annihilation cross-section upper limits obtained using this work and the spectrum calculated by PPPC compared to the Fermi-LAT collaboration analysis \citep{Ackermann:2015zua} for the case of ``standard" DM annihilations into $b\bar{b}$ (top) and $\tau^+\tau^-$ (bottom) final states as a function of the DM mass and assuming a 95\% confidence upper limit set at $\Delta LG(\mathcal{L}) = 2.71/2$. }
\label{fig:compare}
\end{figure}

At present the Fermi-LAT collaboration has produced constraints on the DM annihilation cross-section for DM models that annihilate directly into a few choice final states (e$^+$e$^-$, $\mu^+\mu^-$, $\tau^+\tau^-$, $u\bar{u}$, $b\bar{b}$, W$^+$W$^-$). Since many additional final states (or combinations of these final states) are well motivated, we utilize the Fermi-LAT likelihood profiles for each dwarf galaxy to produce a joint-likelihood analysis which can be analyzed for any $\gamma$-ray spectral shape. Specifically, we utilize the individual likelihood profiles calculated by the Fermi-LAT team for each individual dSph, and then produce an independent likelihood-minimization algorithm to compute $\Delta$LG($\mathcal{L}$) for a joint-likelihood analysis of the entire dwarf galaxy population. We utilize the best-fit J-factors and uncertainties calculated by~\citep{Martinez:2013els}. We note that this J-factor estimation is identical to that employed by the Fermi-LAT team, allowing for a direct comparison between our results and those of~\citep{Ackermann:2015zua}.  As a check of our analysis, we compare our upper bounds with those of \citep{Ackermann:2015zua} assuming dark matter annihilation dominated by a single final state channel. As seen in Figure~\ref{fig:compare}, we obtain results consistent with those obtained by the Fermi-LAT collaboration in the cases of annihilations directly to $b\bar{b}$ and $\tau^+\tau^-$. While we find a small ($<$10\%) offset in the $b\bar{b}$ channel, we note that this is smaller than the deviation stemming from the usage of different particle physics models to characterize the $\gamma$-ray flux from the annihilation to a given dark matter final state.

\begin{figure}[H]
\centering
\includegraphics[scale=0.42,trim=0cm 0cm 0cm 0cm]{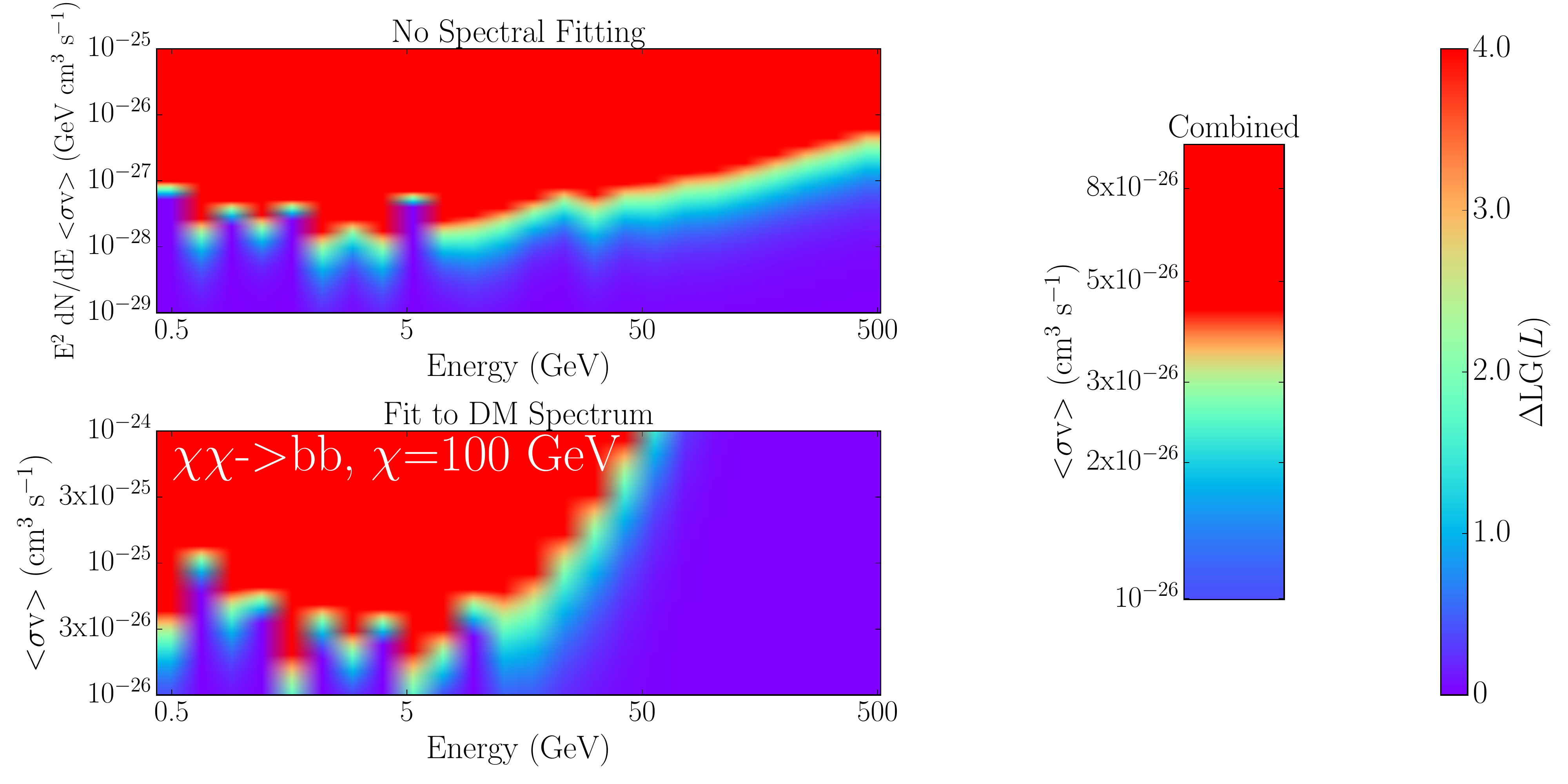} 
\caption{Stages in the analysis of the joint-likelihood limits on the annihilation of particle dark matter. First, the limit is shown in an intermediate state, as a joint-likelihood analysis calculated independently in each spectral bin,with J-factors for each dwarf that are alllowed to float independently in each spectral (top left). Second, a dark matter mass and annihilation pathway is fixed (100 GeV DM annihilating to a $b\bar{b}$ final state) and the J-factor for each dwarf is forced to be consistent over each spectral bin (bottom left). Finally, the combined limits are calculated by summing the LG($\mathcal{L}$) scan in each energy bin to produce a limit on the dark matter cross-section (right).}
\label{fig:BinByBinBounds}
\end{figure}

In Figure~\ref{fig:BinByBinBounds} we demonstrate how the joint-likelihood algorithm is employed to set limits on the annihilation cross-section for a given dark matter mass and annihilation final state. For illustrative purposes, we first (top left) show an intermediate state of the analysis (not directly computed in the analysis routine), where no dark matter spectral information is added to the fitting algorithm, and the J-factor of each dwarf galaxy is allowed to float independently in each energy bin. This provides an upper limit on the particle physics flux ($E^2\frac{dN}{dE} \vev{\sigma v}$) from dark matter annihilation in each energy bin. In this case, the limits are relatively weak, as we are only integrating information from a small energy slice of the Fermi-LAT data. Second (bottom left), we integrate a particle physics model (in this case 100 GeV DM particles annihilating to $b\bar{b}$ final states) and restrict the J-factors in each energy bin to be equivalent. The addition of this particle physics model effectively re-weights each energy bin based on its contribution to the total combined limit. For example, energy bins exceeding 50~GeV become unimportant as the 100~GeV DM particle produces only a negligible $\gamma$-ray flux above this energy. Some bins show small (non statistically significant excesses) which significantly weakens their contribution to the cross-section limit. Finally (right), we show the combined limit on this model, obtained by summing the joint-likelihood analysis in each energy bin shown from Stage 2. In this case, we find a total upper-limit ($\Delta LG(\mathcal{L})$~=~2.71/2) of approximately $\vev{\sigma v}$~=~ 2$\times$10$^{-26}$~cm$^{3}$s$^{-1}$.

\section{Generic Dark Matter Mass Bounds}
Noting that the dark matter annihilation rate scales as $m_\chi^{-2}$, we see from Eq. \ref{eq:gflux} that an upper bound on the gamma-ray flux from DM annihilation can be translated into a lower bound on DM mass, under the assumption of a fixed total dark matter annihilation rate. Here we present lower mass bounds on DM annihilating into multiple final state channels.  This analysis will proceed following the method presented in Ref.~\cite{Carpenter:2015xaa}.  We will assume that DM annihilates into $i$ final state channels, with partial annihilation rates denoted $\vev{{\sigma v}}_{i}$. Specifically, we express the total dark matter annihilation cross-section as a simple sum of partial rates:

\begin{equation}
\vev{\sigma v}_{\text{tot}}=N\vev{\sigma v}_{\text{Th}}= \vev{{\sigma v}}_{1}+\vev{{\sigma v}}_{2}+ \cdots.
\end{equation}

\noindent We have expressed the total DM annihilation rate as some numerical factor times the thermal annihilation cross-section.  $N \vev{\sigma v}_{\text{Th}}$. The first step in this analysis will be to fix the total DM annihilation rate to some desired number. While the total annihilation rate may take any value, the least controversial choice is the thermal rate $(N=1)$, though we note that annihilation rates above or below this level may still be allowable in certain particle-physics models, or within certain non-thermal evolution histories for the universe. One caveat to this assumption is that the total annihilation cross-section to visible channels may be sub-thermal, with an additional annihilation component to invisible final states that ``bleeds" off part of the total annihilation rate.

Defining the partial rates into each channel as in \cite{Carpenter:2015xaa}, $R_i= \vev{{\sigma v}}_{\mathcal{O}_i}/\vev{\sigma v}_{\text{tot}} $, we may divide out by the total rate to get a single constraint
\begin{equation}
1=\Sigma R_i = R_{1}+R_{2}+ R_{3}...
\label{eq:constraint}
\end{equation}
We consider the parameter space of this general analysis to consist simply of the $i$ specific partial DM annihilation rates into each channel.  By fixing the value of the total annihilation rate we now have $i$ \emph{independent} partial annihilation rates and one constraint equation, Eq. \ref{eq:constraint}, reducing our parameter space to $i-1$ parameters.  For three, four, or more annihilation channels we we may then visualize the parameter space as a triangle, tetrahedron, and N-plex as described in \cite{Anandakrishnan:2014pva,Carpenter:2015xaa}. 
 
\begin{figure}[H]
\centering
\includegraphics[scale=0.43]{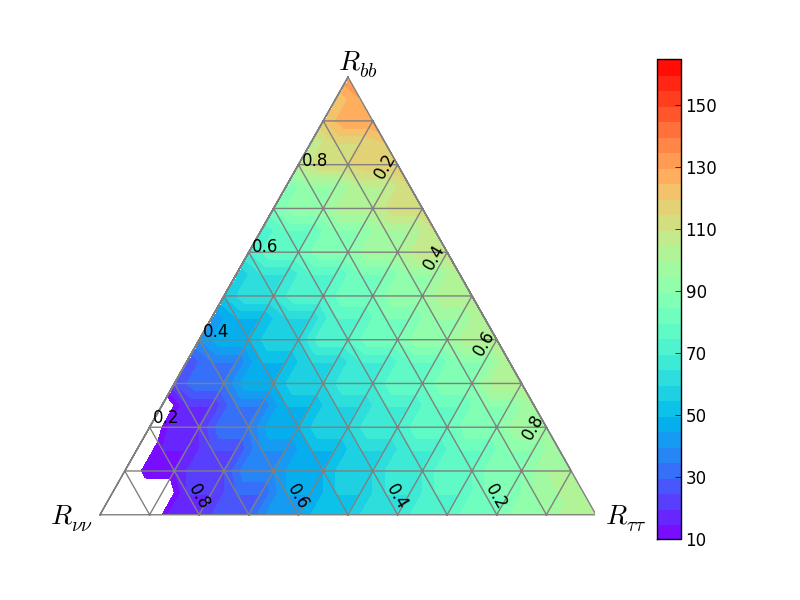}
\includegraphics[scale=0.43]{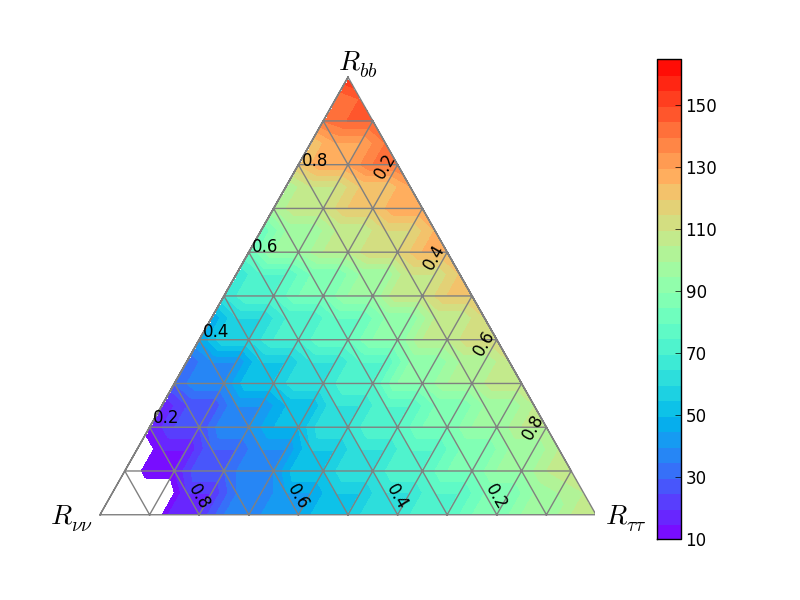}
\caption{The lower bound on the DM mass (GeV) for annihilations into $b\bar{b}$, $\tau^+\tau^-$, and invisible particles for $\vev{\sigma v}_{\text{tot}} = \vev{\sigma v}_{\text{Th}}$.  In the left figure, the photon spectrum is generated with PPPC, while in the right figure, the photon spectrum is generated with DarkSUSY.}
\label{fig:MassBounds2.7}
\end{figure}

We will now consider a generic scenario where we allow DM to annihilate into multiple final state fermions. Here we will limit ourselves to three final state annihilation channels, $b\overline{b}$, $\tau\overline{\tau}$, and neutrinos.  The annihilation into final state neutrinos may be thought of as a generic stand-in for any invisible annihilation channel, which would allow the total DM annihilation rate into visible channels to fall below the thermal annihilation cross-section.  The parameter space is three dimensional. However, once the total annihilation rate is set, we obtain a parameter space that may be represented in two dimensions. Specifically, we can visualize the parameter space by constructing a triangle as in Fig~\ref{fig:MassBounds2.7}.  The points on this triangle specify various admixtures of the partial rates that satisfy the constraint equation above. The vertices of the triangle represent an annihilation rate which is saturated by a single channel. Each outer edge of the triangle corresponds to an annihilation rate which is saturated by only two channels.

Each point on the triangle specifies a different admixture of partial annihilation rates into the $\tau^+\tau^-$, $b\bar{b}$ and invisible channels. Thus, each point specifies a unique $\gamma$-ray spectrum. Having fixed the total DM annihilation rate, we use our our joint-likelihood analysis from the previous sections to place lower limits on the DM mass, as shown in Figure~\ref{fig:MassBounds2.7}. The color contours plotted on the triangle correspond to lower mass bounds on DM at each particular point.  It is clear that the strongest bounds stem from models in which DM annihilates mostly into b-quarks while the weakest bounds result when the invisible channel takes up a significant fraction of the total dark matter annihilation cross-section.  In fact, there is no lower bound when the model allows for greater than 90\% annihilation into invisibles.  We note that in the region where annihilation proceeds only into b's and $\tau$'s, the lower bound on the dark matter masses ranges from 90---150 GeV. 


It should be stressed that the constraints presented in Fig.~\ref{fig:MassBounds2.7} are extremely general. We assume only that the the total dark matter annihilation cross-section is set at the thermal rate, and that the dark matter annihilation cross-section is dominated by annihilations into $b\bar{b}$, $\tau^+\tau^-$, and invisible particles. One may then translate these DM mass bounds, in a less generic way, into various bounds on parameter space of DM models.

\section{EFT completions}
In order to remain agnostic about the high energy details of the DM model, we may describe the interaction between DM and SM fermions as a set of effective operators. This allows for the translation of our lower bounds on the DM mass to bounds on the effective coupling of DM to SM particles. Here, we choose to consider DM annihilating to $b\bar{b}$, $\tau^+\tau^-$ and invisible particles as above. We write a set of dimension 6 operators which couples DM to SM fermions:

\begin{equation}
\mathcal{L}_{\text{f}} =
\frac{1}{\Lambda_b^2}\bar{\chi} \Gamma \chi \bar{b} \Gamma b
+ \frac{1}{\Lambda_{\tau}^2}\bar{\chi}  \Gamma \chi \bar{\tau} \Gamma \tau
+  \frac{1}{\Lambda_{\nu}^2}\bar{\chi} \Gamma \chi \bar{\nu}  \Gamma \nu,
\label{eq:LFermions}
\end{equation}

\noindent where $\Lambda$ is the effective cut-off scale for each operator and  $\Gamma$ specifies if the fermionic currents are scalar, pseudo-scalar, vector, axial vector and so on.  Once the nature of the current is specified, the above model has 4 parameters. There are three effective cut-off scales that specify the DM coupling to each type of fermion, and the fourth parameter is the DM mass $m_{\chi}$.  Each operator in the above equation will contribute to one and only one final state annihilation channel. The total annihilation rate is factorized as a simple sum of the partial rates, which depends on the effective operator couplings with a form proportional to: $a \left(1/\Lambda_b^2 \right)^2+b\left(1/\Lambda_{\tau}^2 \right)^2+c\left(1/\Lambda_{\nu}^2 \right)^2$.
Here we choose to analyze a model with a vector-current,

\be
\mathcal{L}_{\text{f}} =\frac{\kappa_f}{\Lambda_f^2}\chi \gamma^{\mu} \overline{\chi} f \gamma_{\mu} \overline{f}.
\ee

\begin{figure}[H]
\centering
\includegraphics[scale=0.43]{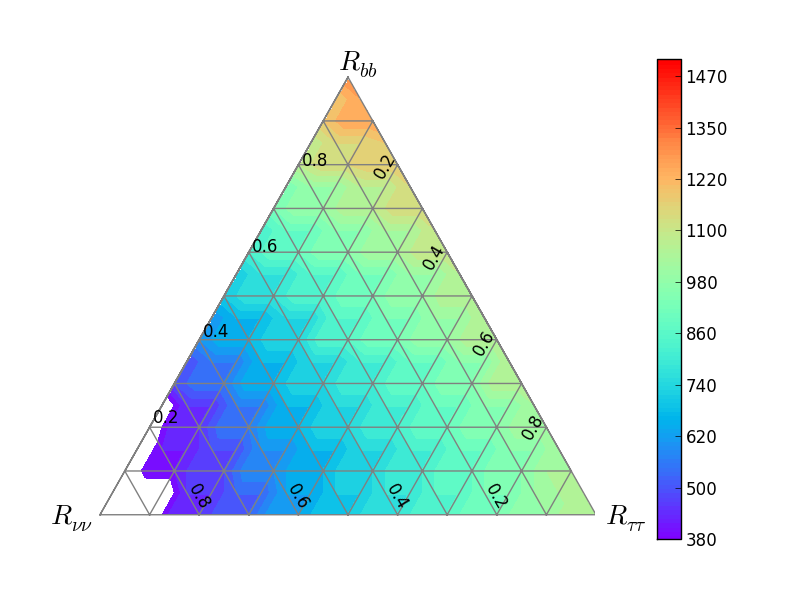}
\includegraphics[scale=0.43]{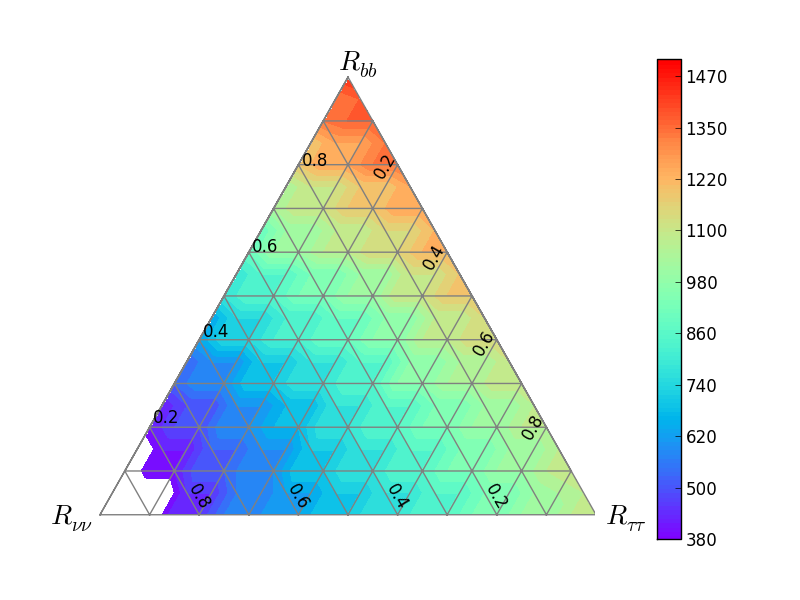}
\caption{The lower bound on the effective coupling (GeV) of DM to b-quarks for annihilations into $b\bar{b}$, $\tau^+\tau^-$, and invisible particles for $\vev{\sigma v}_{\text{tot}} = \vev{\sigma v}_{\text{Th}}$.  In the left figure, the photon spectrum is generated with PPPC. In the right figure, the photon spectrum is generated with DarkSUSY.}
\label{fig:LambdaBBounds2.7}
\end{figure}

\begin{figure}[H]
\centering
\includegraphics[scale=0.43]{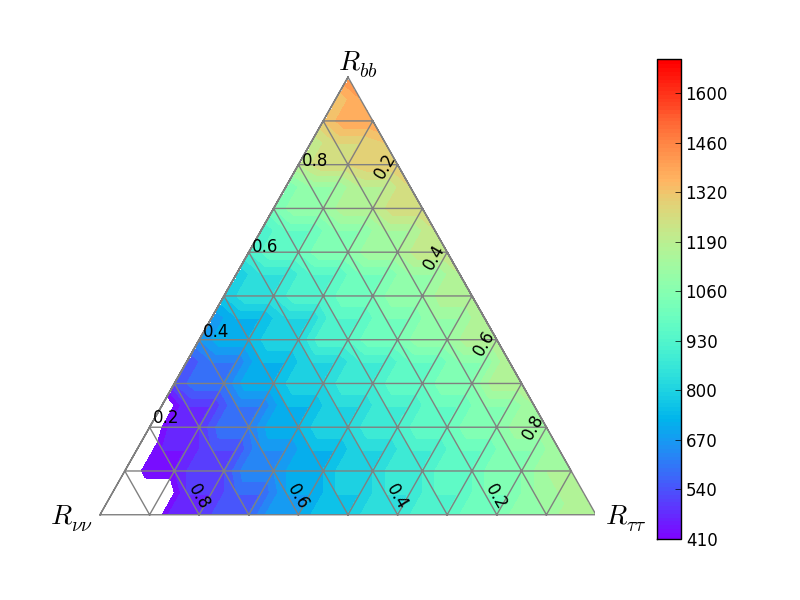}
\includegraphics[scale=0.43]{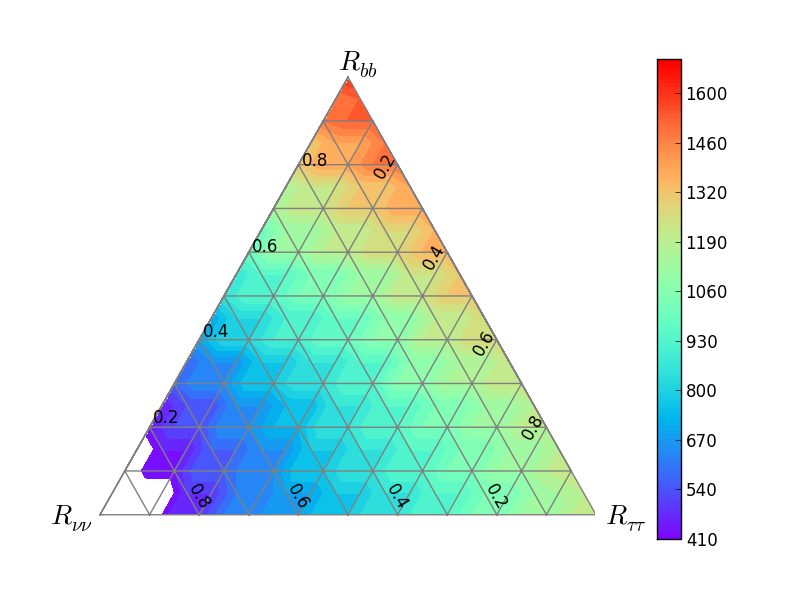}
\caption{The lower bound on the effective coupling (GeV) of DM to $\tau$-leptons for annihilations into $b\bar{b}$, $\tau^+\tau^-$, and invisible particles for $\vev{\sigma v}_{\text{tot}} = \vev{\sigma v}_{\text{Th}}$.  In the left figure, the photon spectrum is generated with PPPC. In the right figure, the photon spectrum is generated with DarkSUSY.}
\label{fig:LambdaTauBounds2.7}
\end{figure}

\noindent With a Lorentz structure specified, we can now translate our generic bounds from the last section into bounds on the effective operator couplings, depicting these limits on an identical triangle morphology.  Again, we fix the total DM annihilation rate to the thermal rate. At each point on the triangle, we utilize the lower-bounds on the DM mass calculated above, which in each case corresponds to a specific value for the effective opperator coefficients. In Figs. \ref{fig:LambdaBBounds2.7} and \ref{fig:LambdaTauBounds2.7} we plot lower bounds on the $\Lambda_b$ and $\Lambda_\tau$ effective operator cut-offs using $\gamma$-ray spectra generated by both PPPC \cite{Cirelli:2010xx} and DarkSUSY~\citep{Gondolo:2004sc}.

Importantly, we note that at each point in the parameter space the lower limit on the operator cutoff is significantly larger than our threshold of twice the dark matter mass ($2 m_{\chi}$). We thus expect that the use of effective operators in this analysis is justified since the effective cutoff significantly exceeds the center of mass energy of the annihilation process. 

\section{Simplified Models}

\begin{figure}[H]
\centering
\includegraphics[scale=0.7,trim=1cm 17.5cm 1cm 2cm]{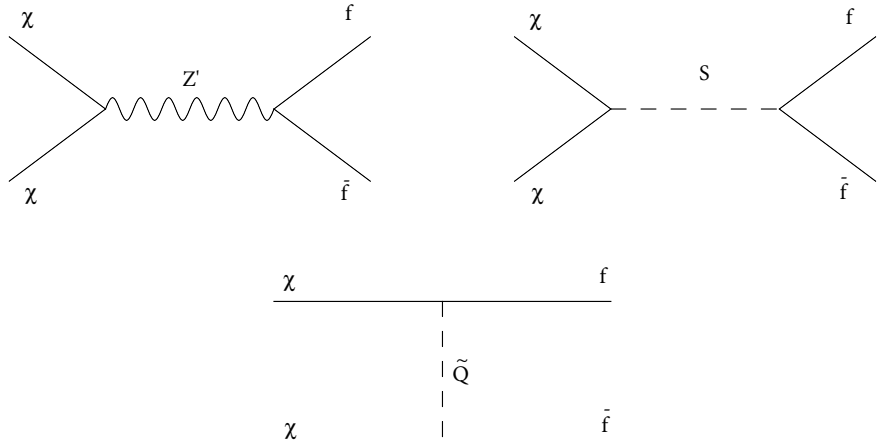}
\caption{Diagrams for DM annihilation in simplified models with a vector mediator (upper left), scalar mediator (upper right), and a t-channel mediator (lower). }
\label{fig:Simp_Mod_Diag}
\end{figure}

We now discuss UV completions for a simple scenario where DM annihilates to SM fermions.  At tree level, the simplest means for DM to talk to the SM is via two basic types of interactions: an s-channel vector mediator, or a t-channel scalar mediator (see Fig. \ref{fig:Simp_Mod_Diag}) \cite{Abdallah:2015ter,DeSimone:2016fbz}.  We show Feynmann diagrams for these simple scenarios below.
In our discussion, we will focus on s-channel vector mediator and t-channel scalar mediator models where DM annihilates to fermionic final states. The case of an s-channel scalar mediator is interesting but more complicated, as the s-channel mediator must mix with the Higgs in order to remain consistent with electroweak symmetry. In this case, the annihilation to final state fermions maybe suppressed in favor of annihilation into Higgses and the couplings to SM fermions will no longer be variable.

\subsection{Vector mediator}

A vector mediated model can be produced in extensions of the SM gauge groups by an extra $U(1)$ that is spontaneously broken in the complete UV theory to obtain a mass for the mediator. We will assume that a UV completion of this model exists that is gauge invariant under new symmetries, anomaly free, and able to generate mass for the mediator. The Lagrangian describing interactions between fermionic DM ($\chi$), the vector mediator ($V_{\mu}$), and the SM fermions ($f_i$) is \cite{Abdallah:2015ter}:

\begin{equation}
\lagr=V_{\mu}\bar{\chi}\gamma^{\mu}(g_{\chi}^{V}-g_{\chi}^{A}\gamma^{5})\chi+\sum_{f}V_{\mu}\bar{f}\gamma^{\mu}(g_{f}^{V}-g_{f}^{A}\gamma^{5})f,
\label{eq:veclagr}
\end{equation}
\\
where the vector ($g_{f}^{V}$) and axial-vector couplings ($g_{f}^{A}$) are assumed to be flavour diagonal.

Using the narrow width approximation, the thermally averaged annihilation cross section to fermionic final states takes on the form:

\begin{equation}
\begin{split}
&\langle\sigma v\rangle (\chi\bar{\chi}\rightarrow V\rightarrow f \bar{f})=
\frac{N^{f}_{c}m_{\chi}^{2}}{2\pi [(M_{V}^{2}-4m_{\chi}^{2})^{2}+\Gamma_{V}^{2}M_{V}^{2}]}\left( 1-\frac{m_{f}^{2}}{m_{\chi}^{2}}\right)^{1/2}\\
&\times
 \left\lbrace
 |g_{\chi}^{V}|^{2}
 \left[|g_{f}^{V}|^{2} \left( 2+\frac{m_{f}^{2}}{m_{\chi}^{2}}\right) +2 |g_{f}^{A}|^{2}\left( 1-\frac{m_{f}^{2}}{m_{\chi}^{2}}\right) \right]
 +|g_{\chi}^{A}|^{2}|g_{f}^{A}|^{2}\frac{m_{f}^{2}}{m_{\chi}^{2}}
\left( 1-\frac{4m_{\chi}^{2}}{M_{V}^{2}}\right)^2
 \right\rbrace,
\end{split}
\end{equation}
\\
where  $N^{f}_{c}=3$ for quarks and $N^{f}_{c}=1$ for leptons, $M_{V}$ is the mass of the vector mediator, and the $m_{f}$ is the mass of the annihilation products. The total width of the vector mediator is:

\begin{equation}
\Gamma_{V}=\frac{M_{V}}{12\pi}\sum_{i} N^{i}_{c}\left(1-\frac{4m_{i}^{2}}{M_{V}^{2}}\right)^{1/2}\left( |g_{i}^{V}|^{2}+|g_{i}^{A}|^{2}+\frac{m_{i}^{2}}{M_{V}^{2}} \left\lbrace 2|g_{i}^{V}|^{2}-4|g_{i}^{A}|^{2} \right\rbrace \right) ,
\end{equation}
\\
where the sum is taken over all kinematically allowed decay modes.
The term in the annihilation rate proportional to $|g_{\chi}^{A}|^{2}|g_{f}^{V}|^{2}$ is p-wave suppressed and has been ignored above.
In order to ensure the validity of the narrow width approximation, one must require $\Gamma_{v} /M_{V}\lsim 0.25$ which corresponds to couplings $g_{i}^{2}\lsim 1$.

In this analysis we will consider the case of pure vector couplings, $g_{i}^{A}=0$, to correspond with the EFT analysis in Section~IV.  The parameter space of the simplified model consists of the DM mass $m_\chi$, the mediator mass $M_V$, and the $i$ mediator-fermion couplings $g_{i}^{V}$.  Using our joint-likelihood upper limits computed in Sections~II and III, we place bounds on the parameter space of this model. In the treatment of this simplified model, unlike the EFT treatment above, we will allow the total DM annihilation rate to float.  We will instead fix the DM mass and also specify the ratios of the partial rates; for example we may consider 100 percent annihilation into $b\bar{b}$, or 33.3 percent annihilation rates into $b\bar{b}$, $\tau^+\tau^-$ and invisible channels respectively.  For each fixed DM mass and ratio of partial rates, we determine the maximal allowed value of the total annihilation rate that saturates the $\gamma$-ray flux bounds.  This then allows us to compute the values of the vector mediator mass and vector couplings which correspond to the total annihilation rate bound.

\begin{figure}[H]
\centering
\includegraphics[scale=0.575]{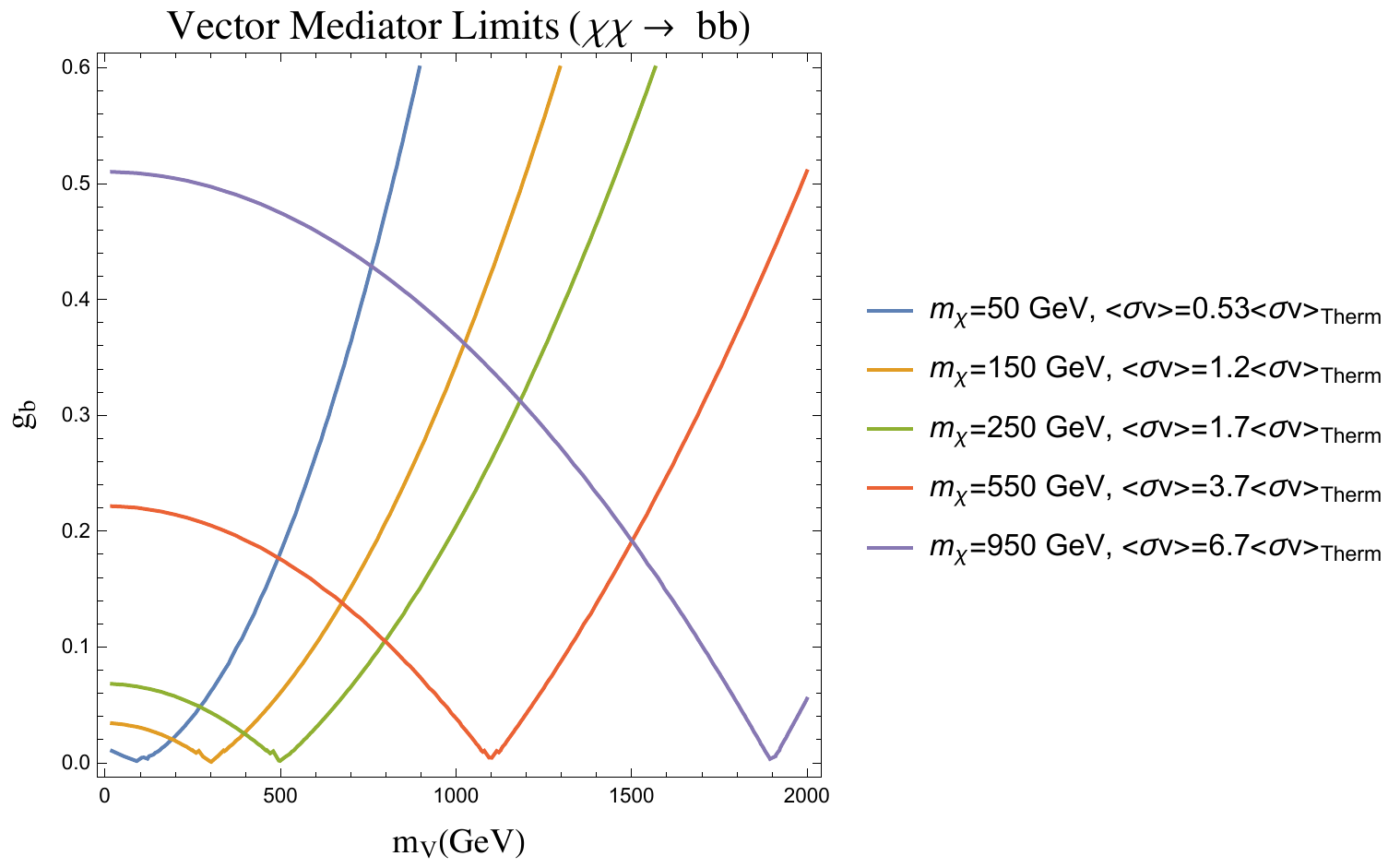}
\includegraphics[scale=0.575]{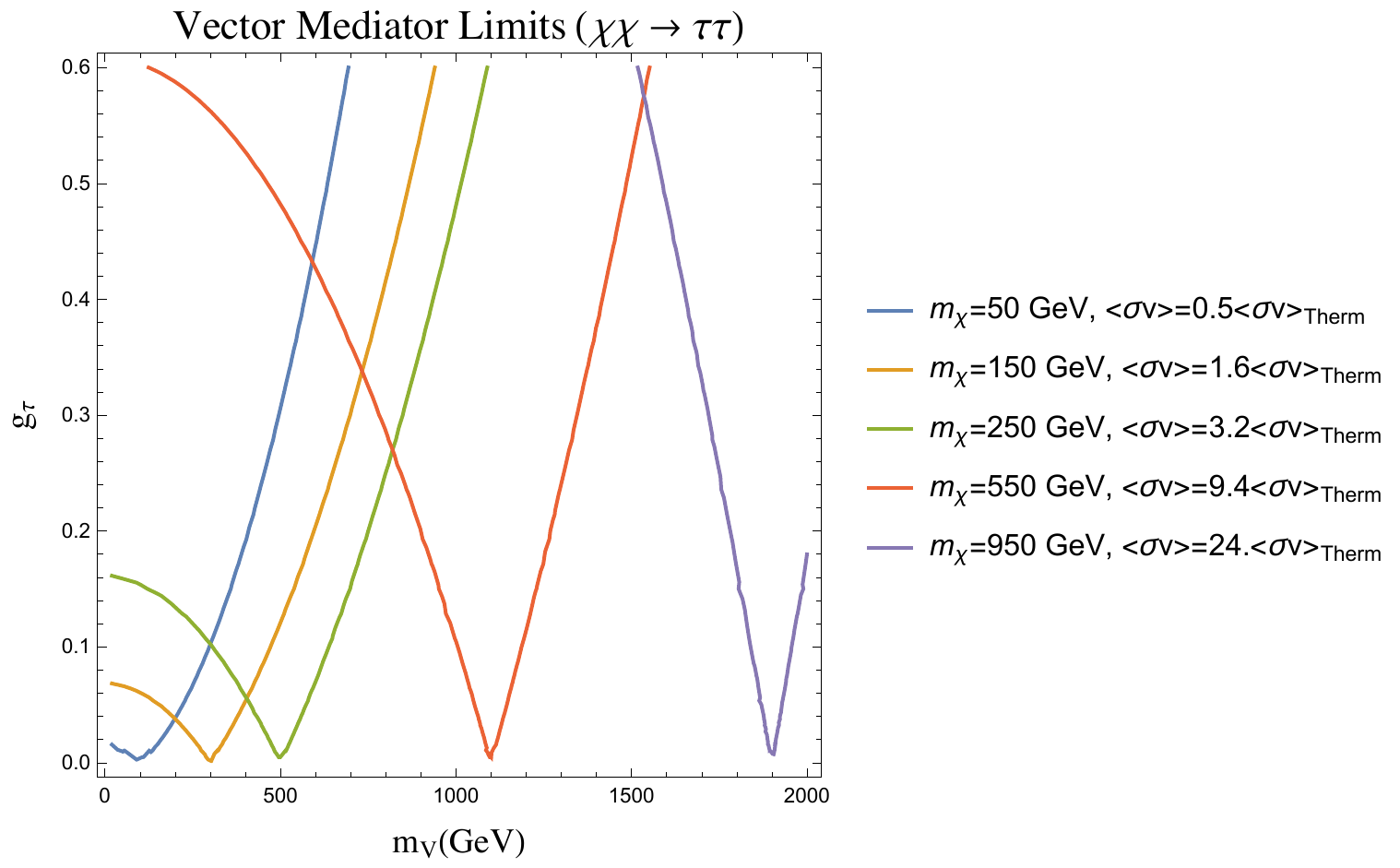}
\includegraphics[scale=0.575]{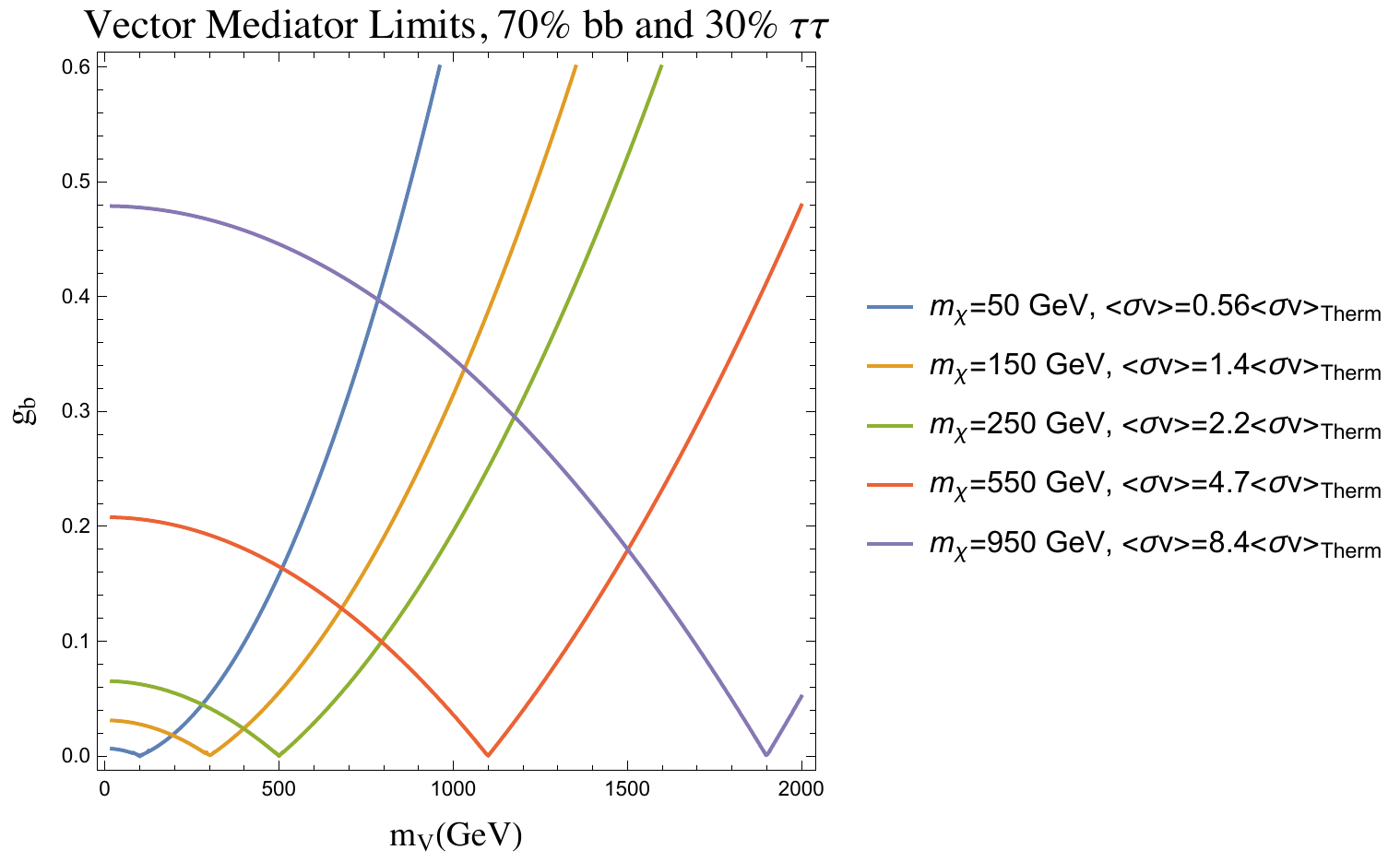}
\includegraphics[scale=0.575]{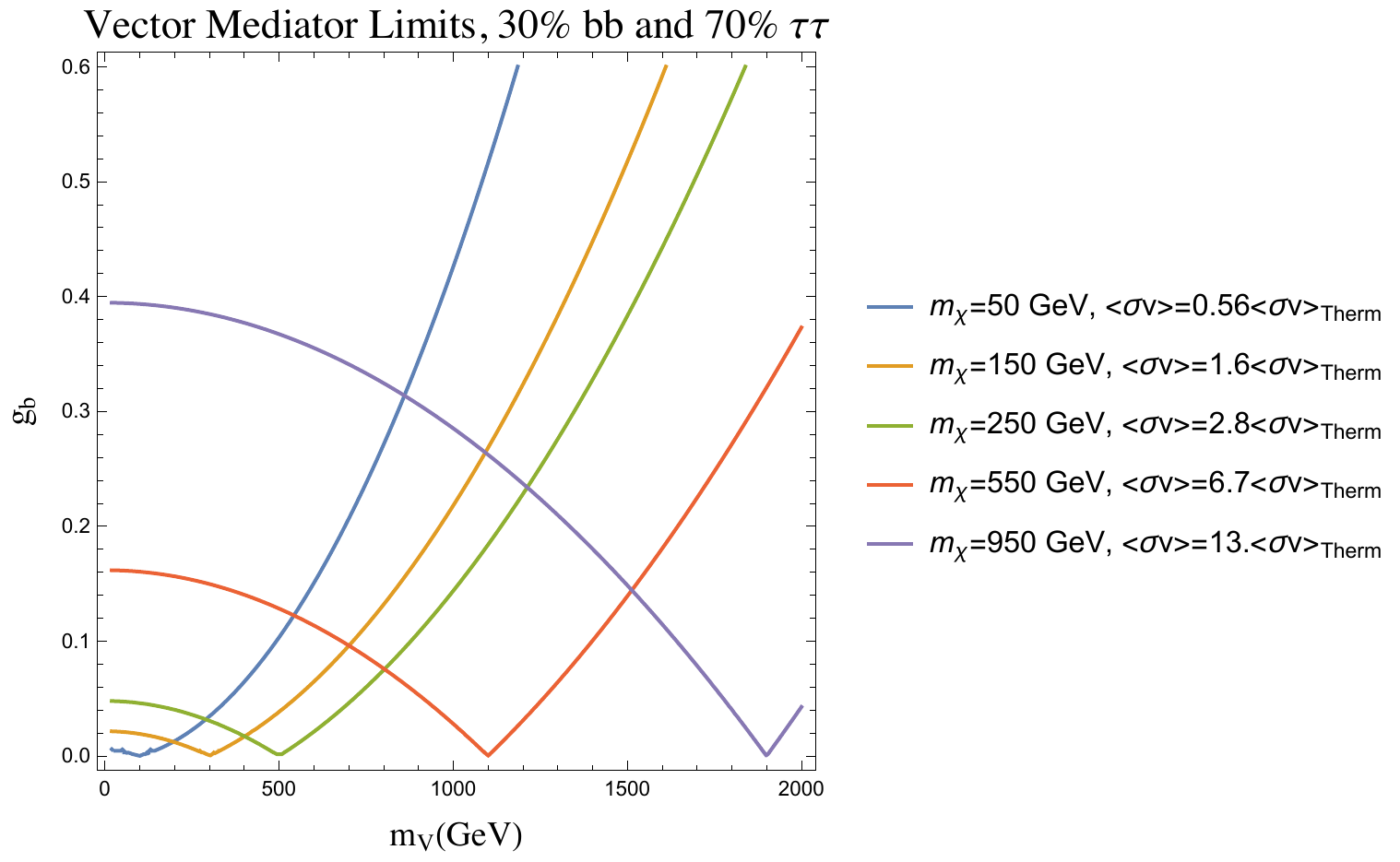}
\caption{Upper bound on the allowed DM-mediator coupling $(g_f)$ as a function of mediator mass for DM annihilation to pure b's (upper left), pure $\tau$'s (upper right), and the 30/70 mixed case (lower) for $g_{\chi}=1$ in the vector mediator simplified model.}
\label{fig:vectorlims}
\end{figure}

In Fig.~\ref{fig:vectorlims} we plot exclusions in the plane of the couplings and mediator mass, for various values of DM mass given the specified admixture of partial annihilation rates.  We plot exclusions for partial rates consisting of: 100\% b-quark - 0\% $\tau$, 0\% b-quark - 100\% $\tau$, 30\% b-quark - 70\% $\tau$, and 70\% b-quark - 30\% $\tau$.  The region above the curves are excluded.  Unsurprisingly we see that parameter space is excluded when the couplings are large, and hence the partial annihilation rate is large. We note another kinematic feature, the parameter space is most restricted for mediator masses $M_{V} \sim 2m_{\chi}$ where the mediating vector particle is produced on resonance.  We see that as the vector mediator is made much heavier than the dark matter mass, the total annihilation cross section drops and the parameter space is less constrained.

\begin{figure}[H]
\centering
\includegraphics[scale=0.69]{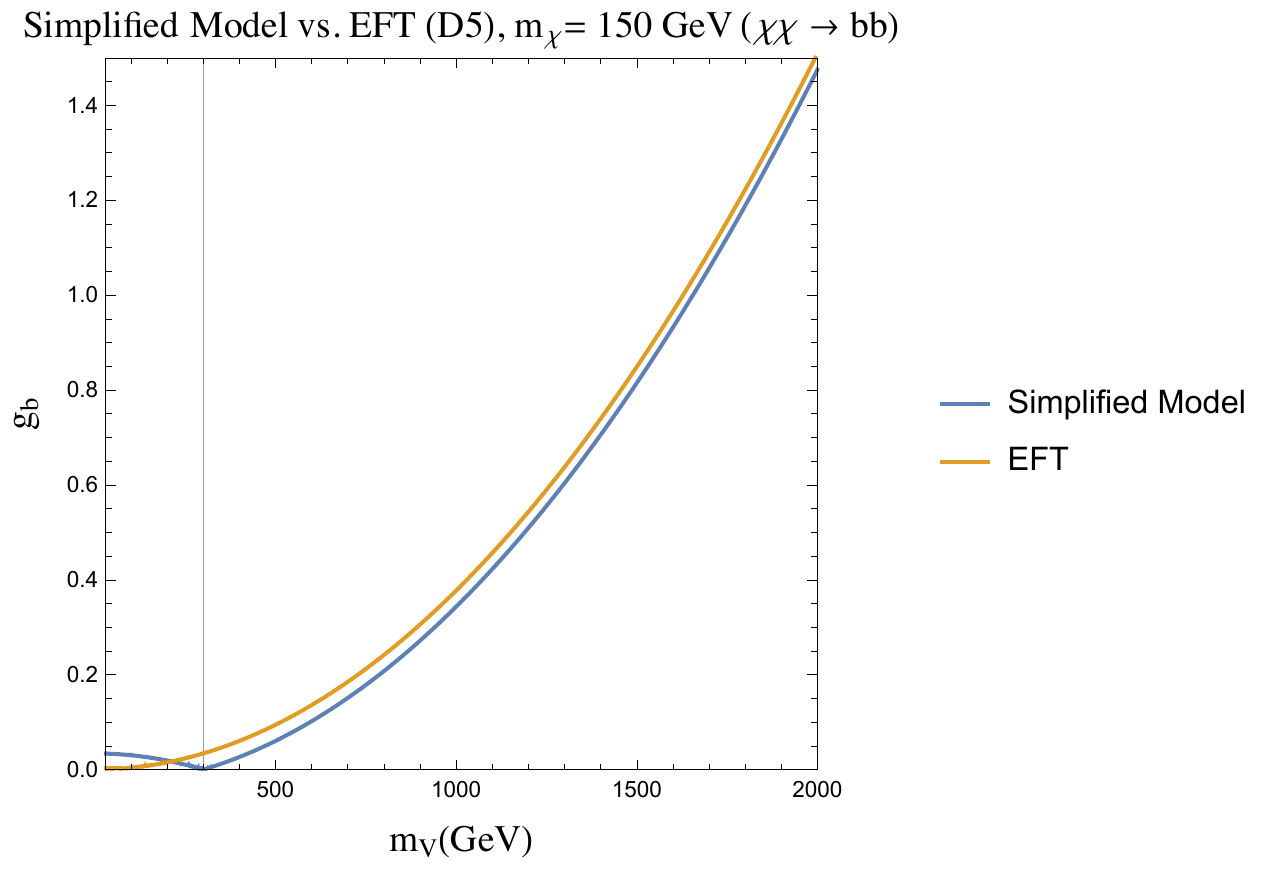}
\includegraphics[scale=0.69]{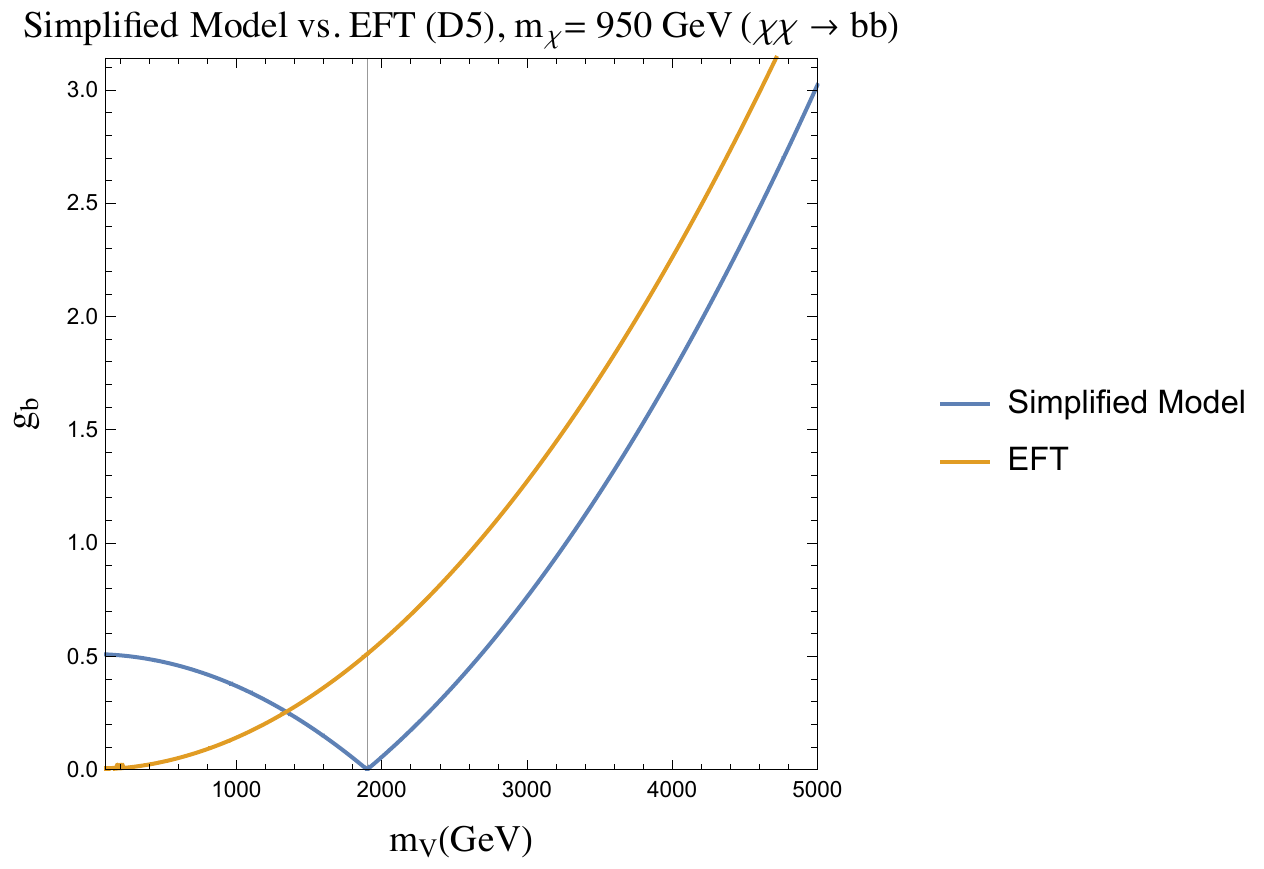}
\caption{Upper bounds on DM-mediator coupling as a function of mediator mass for a vector mediator model compared with assumptions from
effective operator D5 $(\bar{\chi}\gamma^\mu\chi f\gamma_\mu f)$ with $g_{\chi}=1$ and $m_\chi=150$ GeV (left) and $950$ GeV (right).}
\label{fig:EFTSimpComp}
\end{figure}

In Fig.\ref{fig:EFTSimpComp} we compare the exclusion limits from a vector mediator simplified model and effective operator model with the vector current operator $(\bar{\chi}\gamma^\mu\chi f\gamma_\mu f)$.  We have chosen two benchmark masses for the DM particle, one heavy (950 GeV) and one light (150 GeV) and  additionally assume 100\% annihilation into b quarks.  The vertical line indicates where the effective cut-off of the EFT is less than $2 m_{\chi}$.   The mapping from effective coupling to the UV parameters goes as: $\Lambda \sim m_V/\sqrt{g_\chi g_f}$.  In general, we find that the effective theory over-constraints the model. For light DM masses the EFT converges with the simplified model exclusion for cutoffs above approximately 600 GeV. For heavier DM masses the EFT does not closely match the simplified model, and begins to converge only at extremely large coupling values and mediator masses.  It is clear from Fig.\ref{fig:EFTSimpComp} (left), that EFT models with lower DM mass more accurately represent bounds on a more complete model. 

\subsection{T-channel mediator}
In t-channel completions of effective operator models, DM exchanges a massive mediator in a t-channel scattering process to produce two standard model particles.  We take the DM particle to be a fermion, and the mediator to be a scalar. In this case, the mediator must carry SM quantum numbers; the annihilation to quarks will require a colored scalar mediator, while the annihilation to leptons will require a mediator with electroweak quantum numbers. Supersymmetry is the most well known theoretical structure in which to embed the model, but here we will not consider the details of a complete supersymmetric model. The interaction between DM and the right handed SM fermions that we will consider have the form:

\begin{equation}
\lagr=\sum_{i}g_{i}\phi_{i}^{*}\bar{\chi}P_{R}f_{i} + h.c.,
\label{eq:tchannelLag}
\end{equation}
\\
where the sum is over fermions, $P_{R}$ is a right projection operator, and $\phi_i$ are the mediators.
One can also consider coupling to left-handed fields in a similar fashion.
Unlike the vector simplified model above which required only a single vector boson as mediator, here we will require a different mediator for each flavor of Fermion. This will ensure that we do not run aground of flavor violating constraints.  This feature mimics supersymmetric models.

The annihilation cross section corresponding to a t-channel mediator as described in Eq. \ref{eq:tchannelLag} and in the massless final-state limit is \cite{Abdallah:2015ter}:

\begin{equation}
\langle \sigma v \rangle (\bar{\chi}\chi\rightarrow f_{i}\bar{f}_{i})=\frac{N_{c}^{f}g_{i}^{4}m_{\chi}^{2}}{32\pi (M_{i}^{2}+m_{\chi}^{2})^{2}},
\end{equation}
\\
where $M_{i}$ are the the mediator masses, and $N_c$ a color factor.  We note that in the case of light mediators there will be non-trivial bounds on the mediator mass from pair production at colliders. The exact collider bounds will depend on the details of the mediator decay. If $M_{i} > m_{f}+ m_{\chi}$, then the mediator will have a non trivial decay width to a SM fermion plus missing energy.  If this condition is not satisfied the mediator must have additional SM couplings in order to avoid being absolutely stable. The parameter space consists of the DM mass, the $i$ couplings $g_i$ and the $i$ mediator masses $M_i$.
Once these are specified one can calculate all of the specific partial annihilation rates.
As before, the total annihilation rate factorizes into a simple sum of the partial annihilation rates, since the annihilation channels do not interfere; the existence of any one mediator contributes \emph{only} to a single annihilation channel.
We proceed in a similar fashion to the discussion of the vector mediator simplified model.  We will fix the DM mass, and will specify the partial annihilation rate ratios into the final state fermion channels. We may then determine the values of  of $g_i$ and $M_i$ that saturate the total photon-flux bounds.

\begin{figure}[H]
\centering
\includegraphics[scale=0.55]{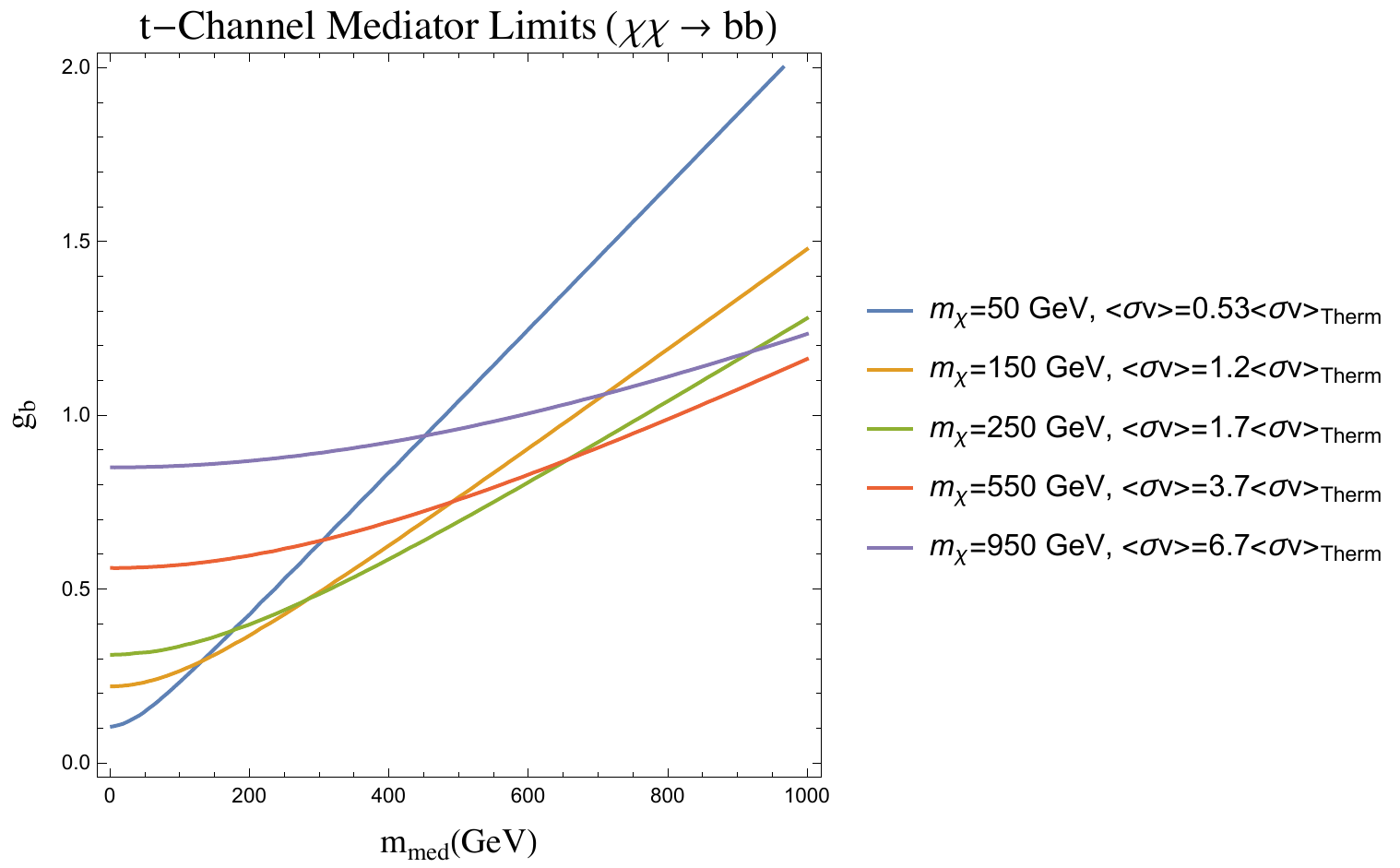}
\includegraphics[scale=0.55]{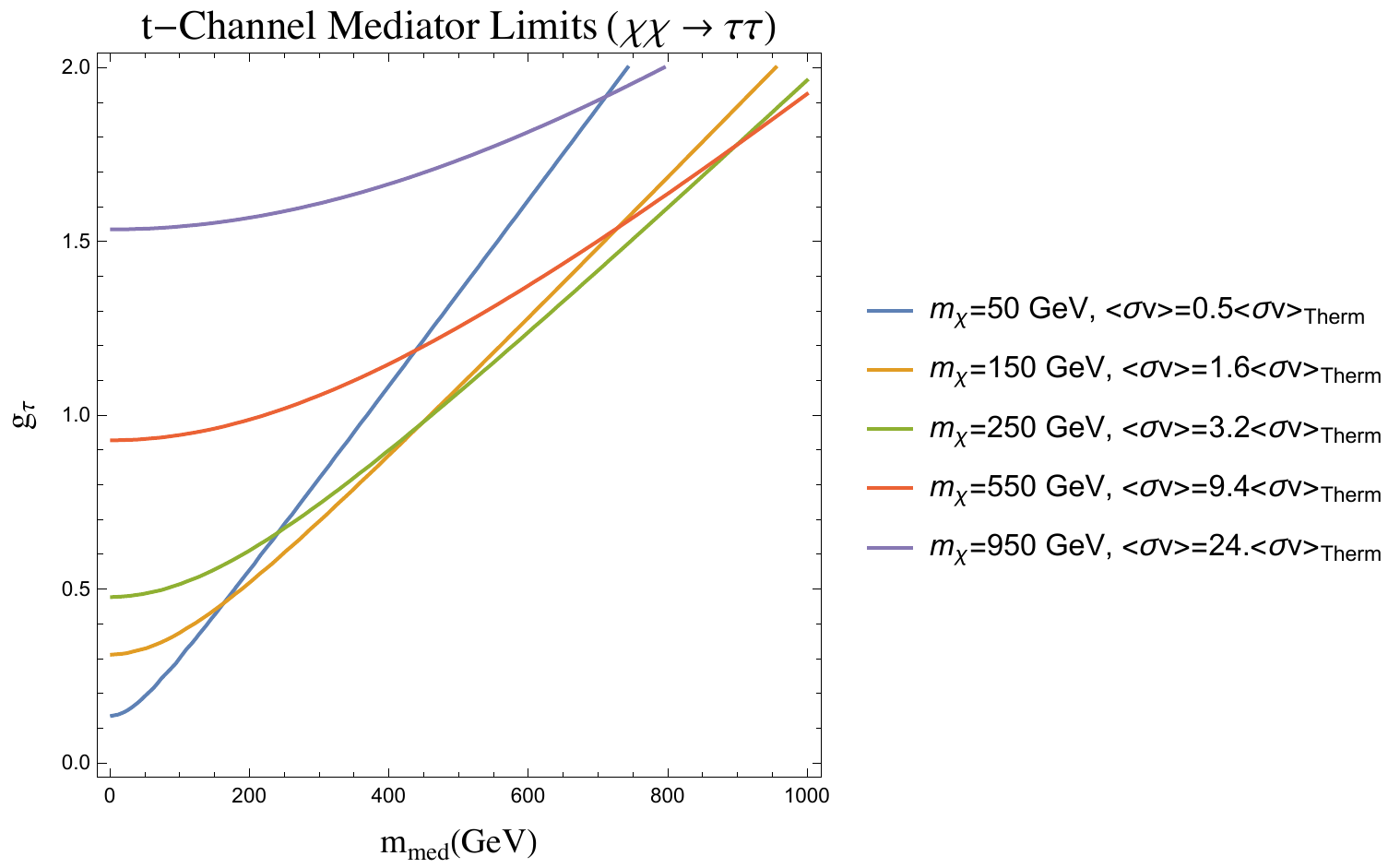}
\caption{Upper bound on the allowed DM coupling to final states bb (left) and $\tau \tau$ (right) in the t-channel mediator simplified model.}
\label{fig:tchanlims}
\end{figure}

\begin{figure}[H]
\centering
\includegraphics[scale=0.55]{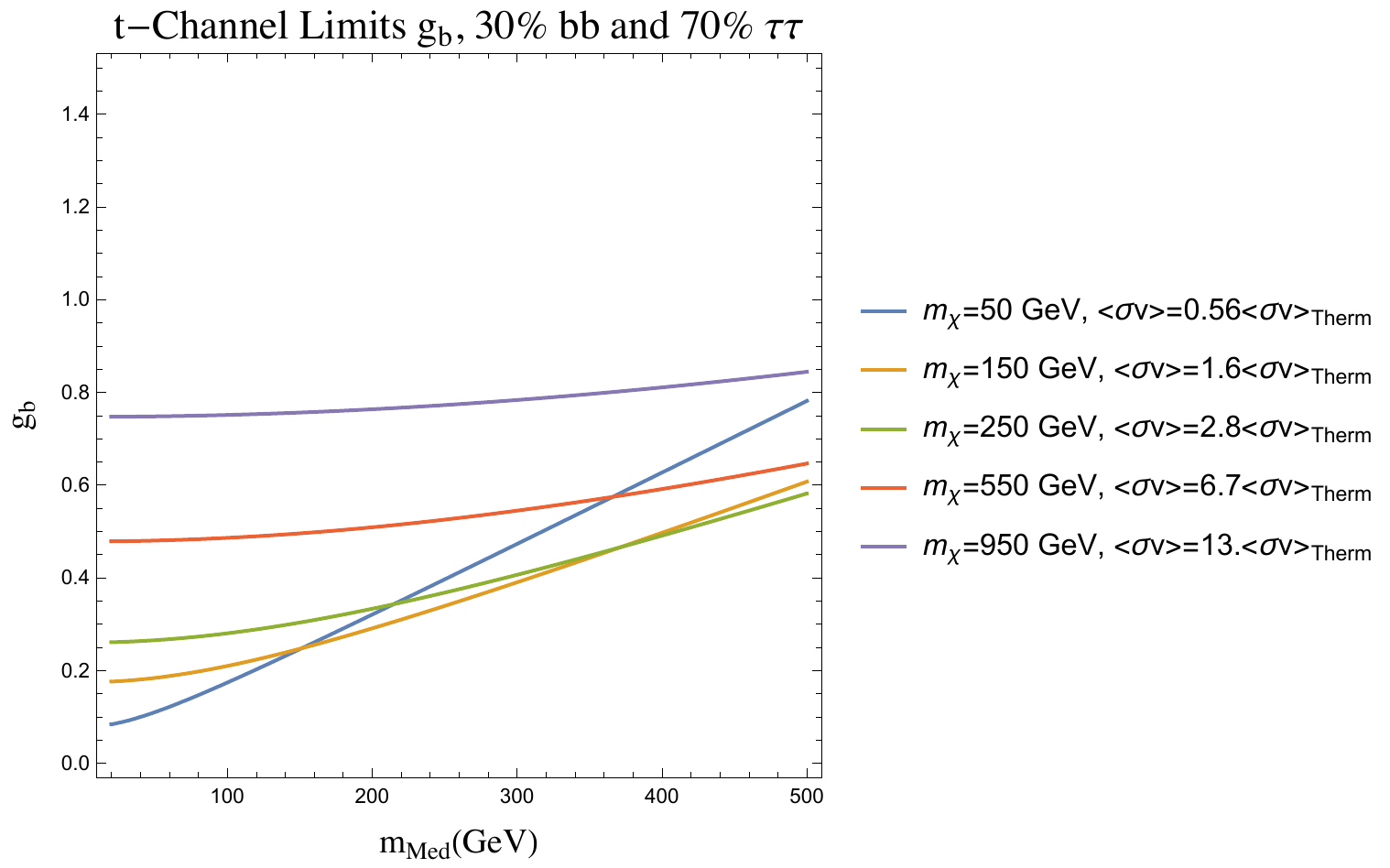}
\includegraphics[scale=0.55]{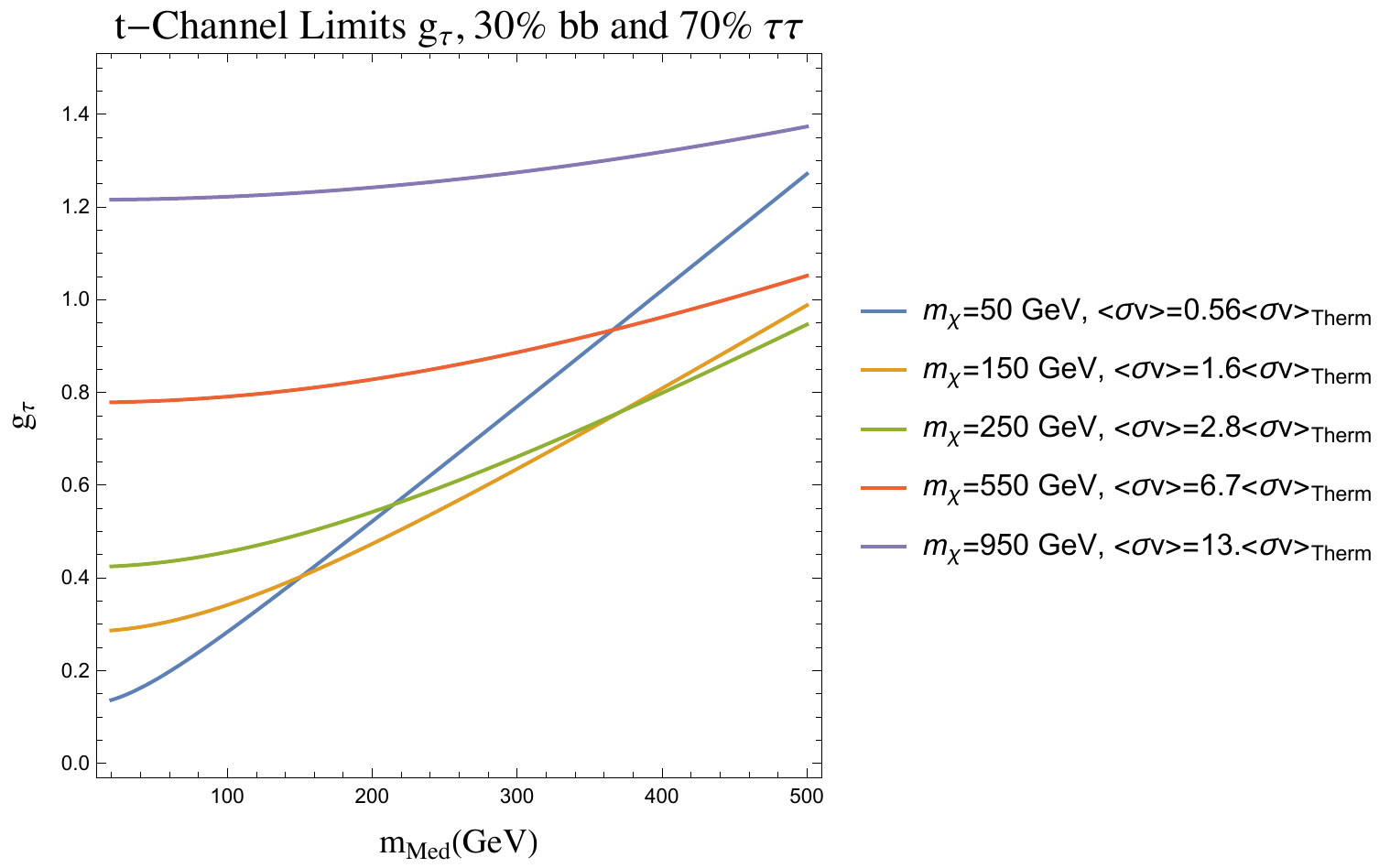}
\caption{Upper bound on the allowed DM coupling vs. mediator mass for a t-channel simplified model where DM annihilates to 70\% $\tau$'s and 30\% b's.  The left plot has the bounds on the parameters related to the b interaction with DM and the right plot is the same for $\tau$ interactions.}
\label{fig:tchanlimsmixed73}
\end{figure}

\begin{figure}[H]
\centering
\includegraphics[scale=0.55]{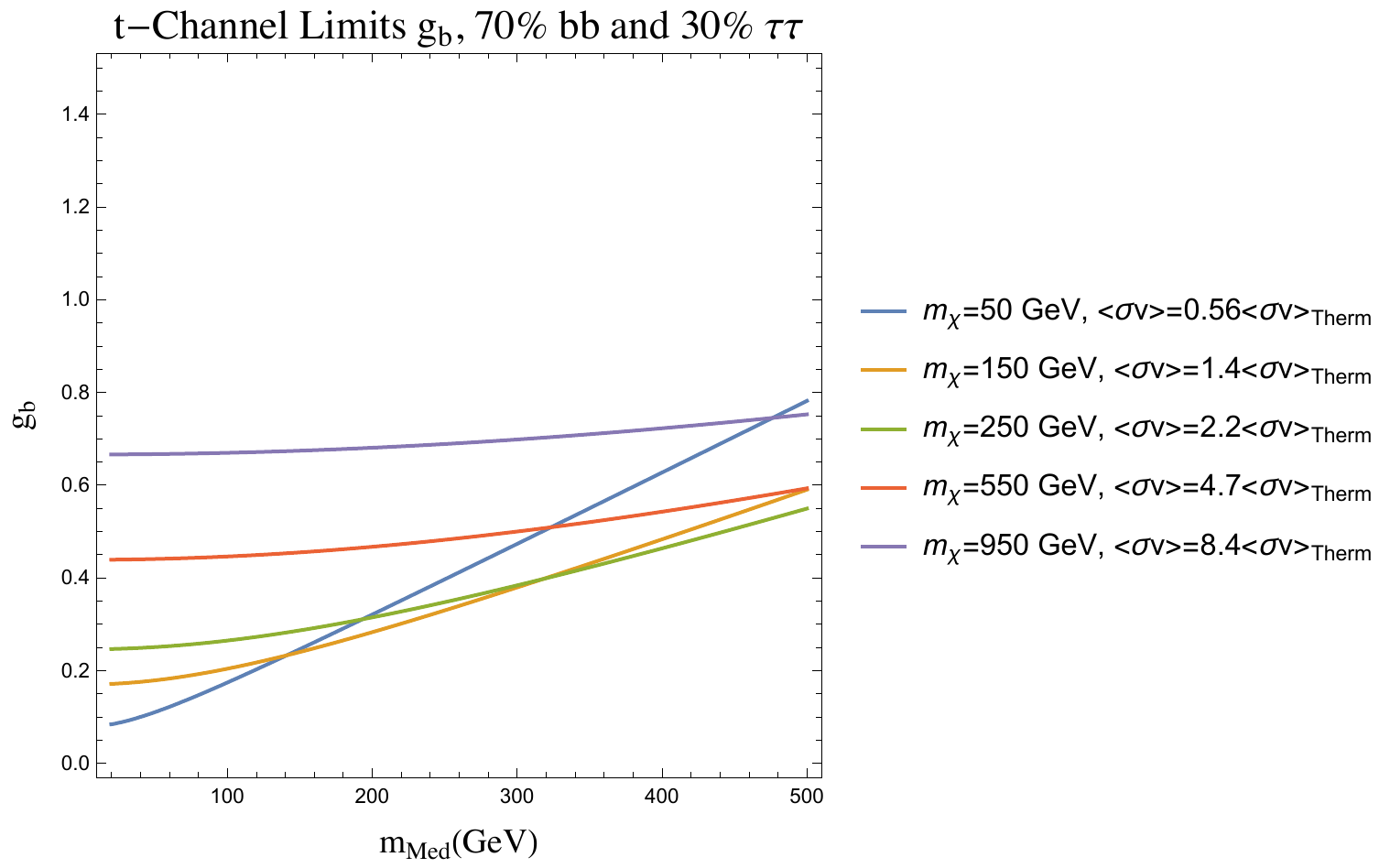}
\includegraphics[scale=0.55]{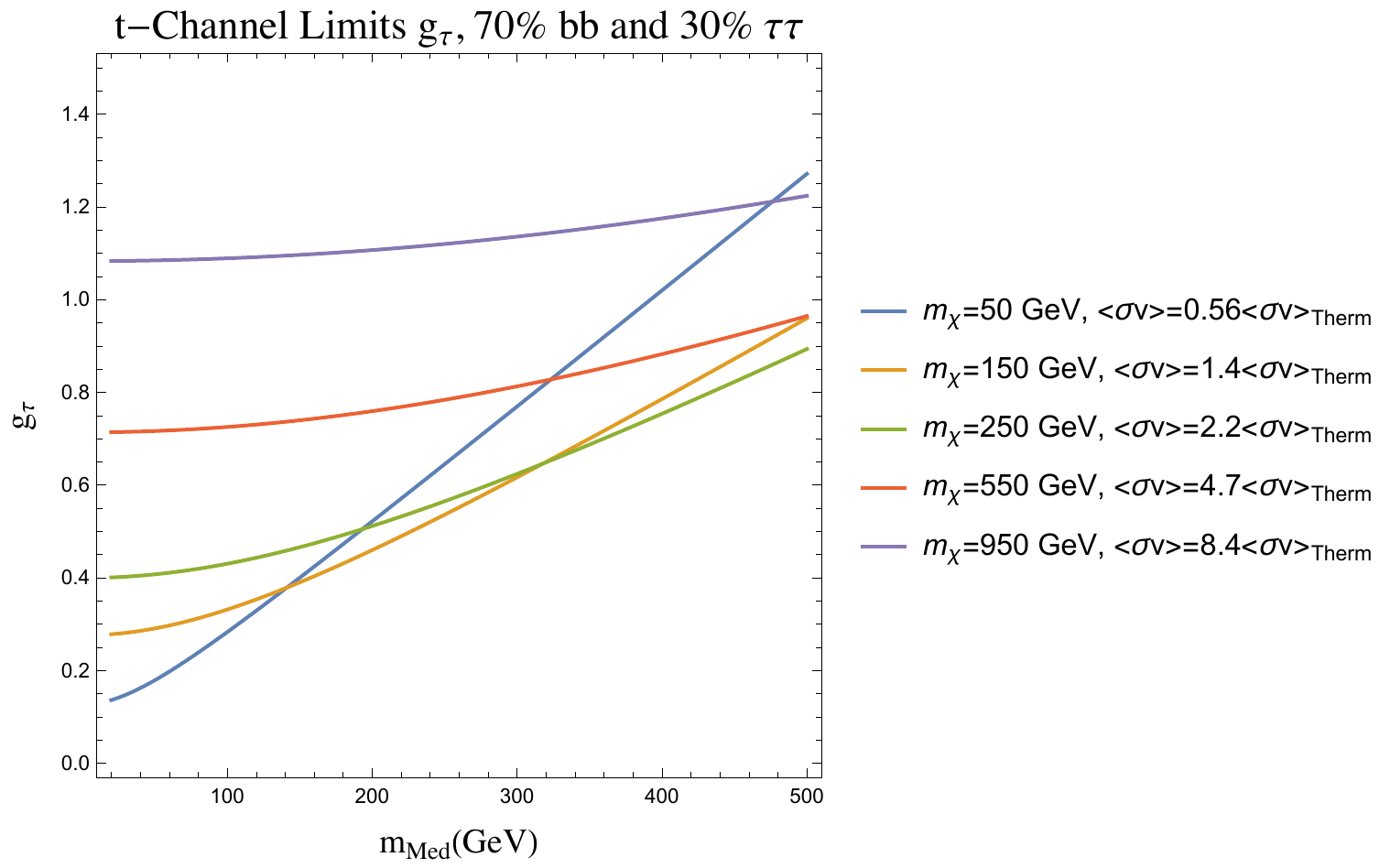}
\caption{Upper bound on the allowed DM coupling vs. mediator mass for a t-channel simplified model where DM annihilates to 30\% $\tau$'s and 70\% b's.  The left plot has the bounds on the parameters related to the b interaction with DM and the right plot is the same for $\tau$ interactions.}
\label{fig:tchanlimsmixed37}
\end{figure}

\begin{figure}[H]
\centering
\includegraphics[scale=0.55]{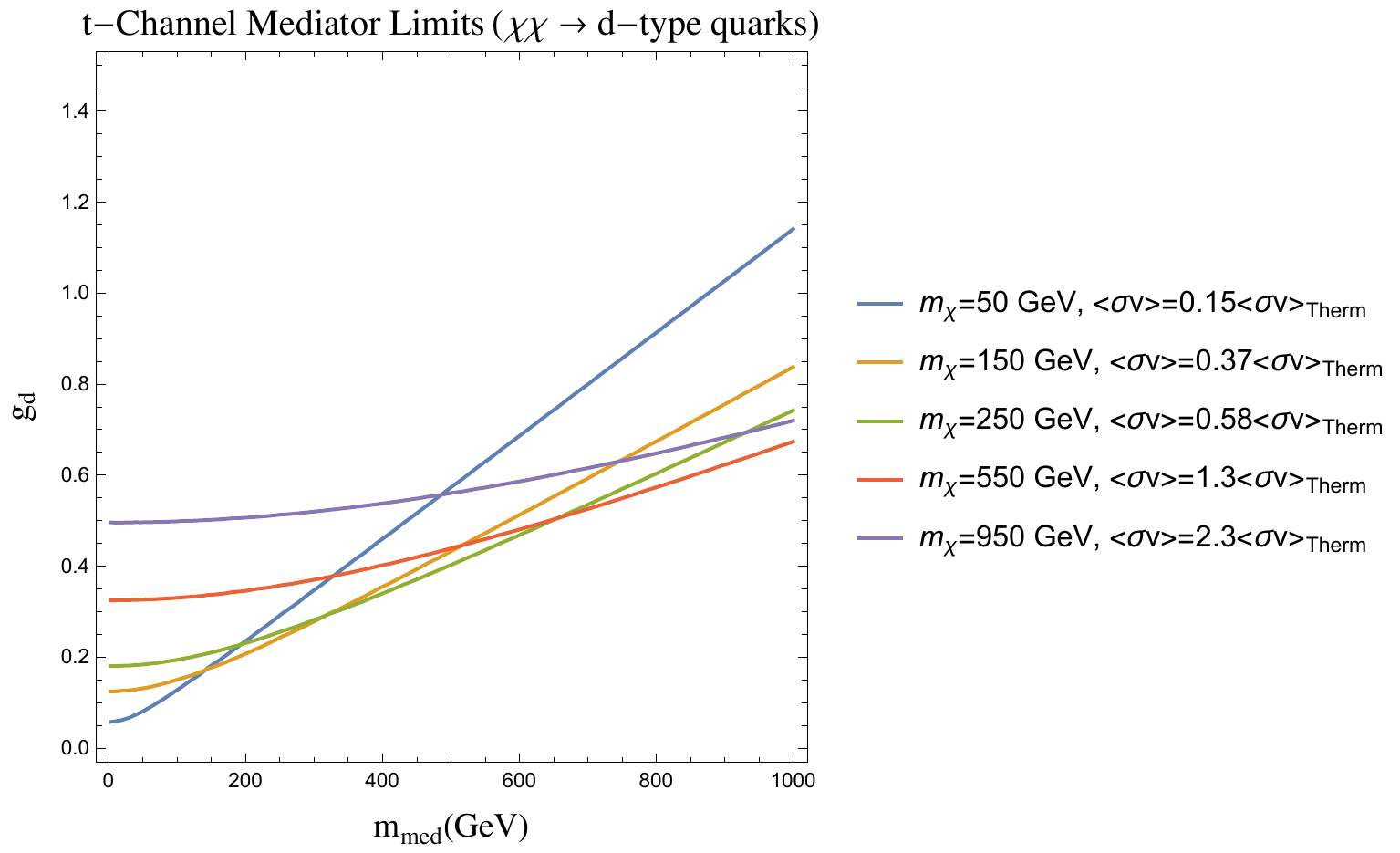}
\includegraphics[scale=0.55]{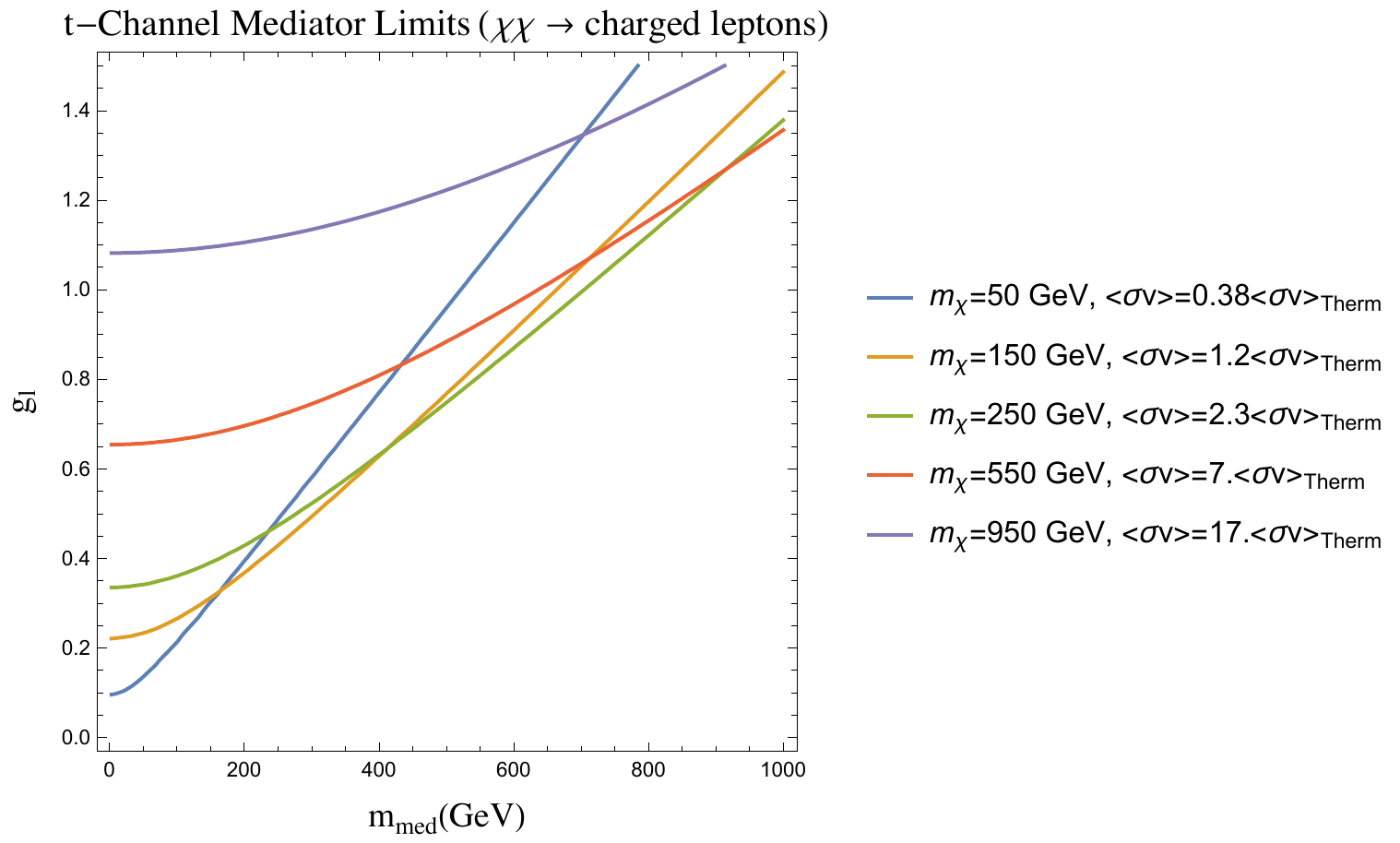}
\caption{Upper bound on the allowed DM coupling to final states of down-type quarks (left) and leptons (right) in the t-channel mediator simplified model.}
\label{fig:tchanlimsfam}
\end{figure}

In Figs. \ref{fig:tchanlims} - \ref{fig:tchanlimsfam} we plot bounds on the  DM model parameter space.  For each curve we have fixed the DM mass, and the bounds are plotted for five different DM masses, $m_\chi = (50, 150, 250, 550, 950)$ GeV.  We also consider various scenarios for the admixture of partial annihilation rates. In Fig. \ref{fig:tchanlims} we consider the scenarios were DM annihilated 100\% into b quarks and 100\% $\tau$'s.  In  Figs. \ref{fig:tchanlimsmixed73} and \ref{fig:tchanlimsmixed37} we consider annihilations to  30\% b's - 70\% $\tau$'s and 70\% b's - 30\% $\tau$'s.   In Fig. \ref{fig:tchanlimsfam} we consider a more flavor democratic scenario where we consider annihilation into 30\% down-type quarks - 70\%charged leptons and 70\% down-type quarks - 30\% charged leptons respectively.

We plot exclusions in the coupling vs mediator mass plane. The region above the curves is excluded. As in the previous simplified model analysis, we see that regions with large couplings have large annihilation cross section and are thus excluded.  We also note that when the DM mass is light, the DM annihilation cross section scales like $\sim m_{\chi}^2/ M_i^4$, therefore the total photon flux drops sharply with the fourth power of the mediator.  Therefore we see that when DM masses are light, our limits considerably weaken as the mediator mass increases.

\section{Conclusions}

By utilizing data from the Fermi-LAT Collaboration's joint-likelihood analysis of the dwarf spheroidal galaxies of the Milky Way~\citep{Ackermann:2015zua}, we have produced stringent, and generic, limits on the dark matter annihilation cross-section in models where DM annihilates to multiple final state channels. Specifically, we have produced new joint-likelihood analyses of 15 dSphs with relatively well-measured J-factors and calculated the lower-limits on the dark matter mass for models which annihilate at the thermal cross-section to arbitrary admixtures of $b\bar{b}$, $\tau^+\tau^-$ and invisible final states. We note that the tools presented here could be generalized to produce limits on the dark matter annihilation rate assuming any dark matter mass and $\gamma$-ray spectrum. 

We have analyzed the resulting limits in two ways. First, we have presented the constraints as lower limits on the cutoff-scales in an effective field theory approach. We find that this approach is justified as the lower-limits on the cutoff-scales typically significantly exceeds the center of mass energy for the dark matter annihilation event. We have additionally explored simplified models as completions of the EFT scenario. Specifically, we have considered a vector mediated s-channel completion as well as a scaler-mediated t-channel completion. We have presented constraints for each model in the parameter space of the mediator mass vs. the mediator-SM coupling, and have considered flavor restricted scenarios where DM couples only to b-quarks and $\tau$'s, as well as a flavor-democratic annihilation scenario. In order to test the validity of the effective operator paradigm we compared the results from our EFT scenario against those derived for the vector mediated simplified models.  We find that for mediator masses above a few times $2 m_{\chi}$ the results matched fairly well. However the EFT severely over-constrained the cut-offs for models close to the resonance production limit.




There are significant extensions which could be followed along these lines. We note that the tools presented here could be generalized to produce limits on the dark matter annihilation rate assuming any dark matter mass and $\gamma$-ray spectrum. Of particular interest would be models where DM annihilates into electroweak gauge bosons and photons. These models are interesting because they generally require a loop level process in order to couple DM to the SM. Another very fruitful avenue would be to interpret our stacked constraints in the supersymmetric paradigm of the PMSSM.

\section{Acknowledgements}
This work was made possible with funds from DOE grant DE-SC0013529. TL is supported by the National Aeronautics and Space Administration through Einstein Postdoctoral Fellowship Award No. PF3-140110.

\bibliography{Triangles.bib}

\end{document}